\documentclass[hidelinks,onefignum,onetabnum]{siamart251104}

\usepackage{amsfonts}
\usepackage{graphicx}
\usepackage{wrapfig}
\usepackage{enumitem}
\usepackage{array}
\usepackage{tabularx}
\ifpdf
  \DeclareGraphicsExtensions{.eps,.pdf,.png,.jpg}
\else
  \DeclareGraphicsExtensions{.eps}
\fi

\usepackage{amsopn}
\usepackage{xcolor}

\makeatletter
\newcommand{\manualtablecaption}[2]{%
  \refstepcounter{table}\label{#1}%
  \global\@capstartfalse
  {\small\centering\textsc{Table~\thetable}. \textit{#2}\par}%
  \vspace{0.5\baselineskip}%
}
\makeatother
\usepackage{url}
\IfFileExists{pythonhighlight.sty}{\usepackage{pythonhighlight}}{}
\newtheorem{prop}{Proposition}
\newcommand{\Tr}{\operatorname{Tr}}
\ifpdf
\hypersetup{
  pdftitle={A Human-AI Collaborative Workflow for Mathematical Discovery: A Case Study in Grover-Compatible Riemannian Optimization},
  pdfauthor={}
}
\fi

\newcommand{\EnableCompactMode}{%
    \setlist{nosep}
    \setlist[enumerate]{label=\arabic*., leftmargin=*, wide=0pt}
    \setlist[itemize]{label=\textbullet, leftmargin=1.2em}
}

\begin{document}
\raggedbottom %

\headers{Human-AI Workflow for Mathematical Discovery}{C. Li, Z. Lai, D. An, J. Hu and Z. Wen}

\title{{A Human-AI Collaborative Workflow for Mathematical Discovery: A Case Study in Grover-Compatible Riemannian Optimization}}

\author{Chenyi Li\thanks{Beijing International Center for Mathematical Research, Peking University, Beijing, 100871, People's Republic of China (\email{lichenyi@stu.pku.edu.cn,
lai\_zhijian@pku.edu.cn,
dongan@pku.edu.cn,
wenzw@pku.edu.cn}).
}
\and
Zhijian Lai\footnotemark[1]
\and
Dong An\footnotemark[1]
\and
Jiang Hu\thanks{Yau Mathematical Sciences Center,
Tsinghua University, Beijing, 100190, People's Republic of China (\email{jianghu@tsinghua.edu.cn}).
}
\and
Zaiwen Wen\footnotemark[1]}

\maketitle
\begin{abstract}
We investigate how large language models can be used as research tools in scientific computing while preserving mathematical rigor. We propose a human-in-the-loop workflow for interactive theorem proving and discovery with LLMs. Human experts retain control over problem formulation and assumptions, while the model searches for proofs or contradictions, proposes candidate properties and theorems, and helps construct structures and parameters that satisfy explicit constraints, supported by numerical experiments and simple verification checks. Experts treat these outputs as raw material, further refine them, and organize the results into precise statements and rigorous proofs. We instantiate this workflow in a main case study on the connection between manifold optimization and Grover's quantum search algorithm, where the pipeline identifies invariant subspaces and explores Grover-compatible retractions. The main case study uses the corresponding Grover-compatible convergence analysis, including an \(O(\sqrt N\log(1/\varepsilon))\) PL-based bound established in the companion mathematical work, to illustrate the refinement stage of the workflow. Prompt records and reusable templates for implementing the workflow are provided. We further include a multi-oracle case study, document representative failed and corrected routes arising from this setting, and provide a structured failure-mode analysis.
\end{abstract}



\section{Introduction}\label{sec:intro}

Recent advances in artificial intelligence, especially large language models, show strong capabilities in analyzing long text \cite{zhang2020pegasus}, consolidating domain knowledge \cite{lewis2019bart}, and supporting literature-grounded querying through retrieval-augmented pipelines \cite{lewis2020retrieval}. They also achieve steady gains in stepwise mathematical reasoning \cite{lewkowycz2022solving,wei2022chain}. At the same time, mathematical research remains a demanding enterprise that requires sustained abstraction, coherence over long arguments, and verifiable correctness \cite{thurston2006proof,tao2007good}. In practice, the process is conducted by human experts and often iterative. Human experts pose questions, build candidate structures, test them on examples or with tools, and keep what survives as statements with reusable arguments \cite{lakatos2015proofs,polya1957solve}. Motivated by these developments, we study how LLMs can support nontrivial mathematical research while keeping human expertise central. We discuss several examples of human-AI interaction in mathematical theorem proving and theorem discovery. Model outputs are tested by calculations or numerical experiments. Failed routes are rejected, and repaired and verified ideas are promoted to theorem statements and proofs.

A growing body of literature utilizes machine learning to assist mathematical research and to tackle difficult problems across algebra, geometry, and related areas. In algebra, collaborations between mathematicians and AI reveal useful relations in representation theory and guide the formulation and proof of new results \cite{davies2021advancing}. Recent studies use data-driven methods to rediscover classical algebraic inequalities \cite{dong2024machine}. In algebraic combinatorics, curated datasets focus on the conjecturing process and support machine-assisted exploration of patterns that lead to precise and testable statements \cite{chau2025machine}. In geometry, AI systems have produced human-readable proofs for challenging Euclidean problems and strong performance on Olympiad-style benchmarks by combining search, deduction, and verification \cite{trinh2024solving,chervonyi2025gold}. Data-driven analyses of Calabi-Yau spaces reveal structural regularities in algebraic geometry and motivate conjectures that can be vetted by experts \cite{he2024distinguishing}. Beyond individual domains, Ramanujan Machine procedures generate candidate identities for fundamental constants and invite human checks and proof \cite{raayoni2021generating}. Symbolic regression and equation discovery pipelines recover concise laws from data and suggest invariants and lemmas for further study \cite{udrescu2020ai,brunton2016discovering}. Most recently, OpenAI reported ten advances on long-standing problems in mathematics and theoretical computer science through an AI--human workflow in which an AI model generated candidate arguments and humans helped prepare them as mathematical manuscripts \cite{openai2026tenadvances}.
These works promote automatic workflows for theorem discovery and proving, in which models propose, rank, and test candidate statements, while human researchers shape definitions, control scope, and turn promising leads into theorems.

A separate line of research targets solutions to mathematical competitions, especially IMO problems. Specialized domain models such as DeepSeekMath-V2 achieve gold-medal performance on IMO tasks via purely natural-language proofs with substantial test-time computation and dedicated self-verification \cite{shao2025deepseekmathv2}. In parallel, frontier general-purpose reasoning models from Google reach comparable scores on IMO 2025 style problems. Building on these backbones, Huang and Yang design a model-agnostic verification and refinement pipeline that attains gold-medal performance at IMO 2025 by systematically generating, checking, and refining candidate solutions \cite{huang2025gemini}. Complementing this natural-language line, DeepMind's AlphaProof system combines reinforcement learning with formal proof search in Lean and achieves silver-medal performance on IMO 2024 problems under fully formal verification \cite{hubert2025olympiad}. These advances suggest that natural language and formal pipelines can both approach top human performance on Olympiad benchmarks. Formal verification provides a different reliability standard. Lean, Coq, and Isabelle encode definitions and proofs in machine-checkable languages \cite{deMoura2015Lean,bertot2013interactive,nipkow2002isabelle}, and recent LLM integrations improve proof-script search \cite{yang2023leandojo,polu2020generative,ren2025deepseek}.

Beyond theorem proving, many recent efforts use AI to assist scientific research, with most systems targeting applied areas such as biomedicine \cite{huang2025biomni}, the natural sciences \cite{yang2023ai}, and machine learning \cite{li2024mlr}. Deep research systems \cite{huang2025deep, xu2025comprehensive} go further by producing end-to-end reports and maintaining plans over long horizons, while sustained human supervision remains important for strong conclusions. A parallel line builds agentic systems for automating parts of the research workflow, ranging from specific agents to AI scientist frameworks and research copilots, showing feasibility in structured settings \cite{lu2024ai,gandhi2025researchcodeagent}. Automated survey methods can review the relevant background literature \cite{wang2024autosurvey,liang2025surveyx}. For methodology implementation, recent systems pursue automated code generation for research, including project-level context refinement with compiler feedback \cite{bi2024iterative}. In parallel, automated academic-writing tools target reference handling, citation integrity, and drafting support \cite{sun2024metawriter,fuad2024llm}. Finally, automated review systems focus on argumentation quality and fairness, while mathematical rigor still requires expert verification \cite{liu2023reviewergpt,wang2020reviewrobot,idahl2024openreviewer}.

\begin{figure}
    \centering
    \includegraphics[width=1\linewidth]{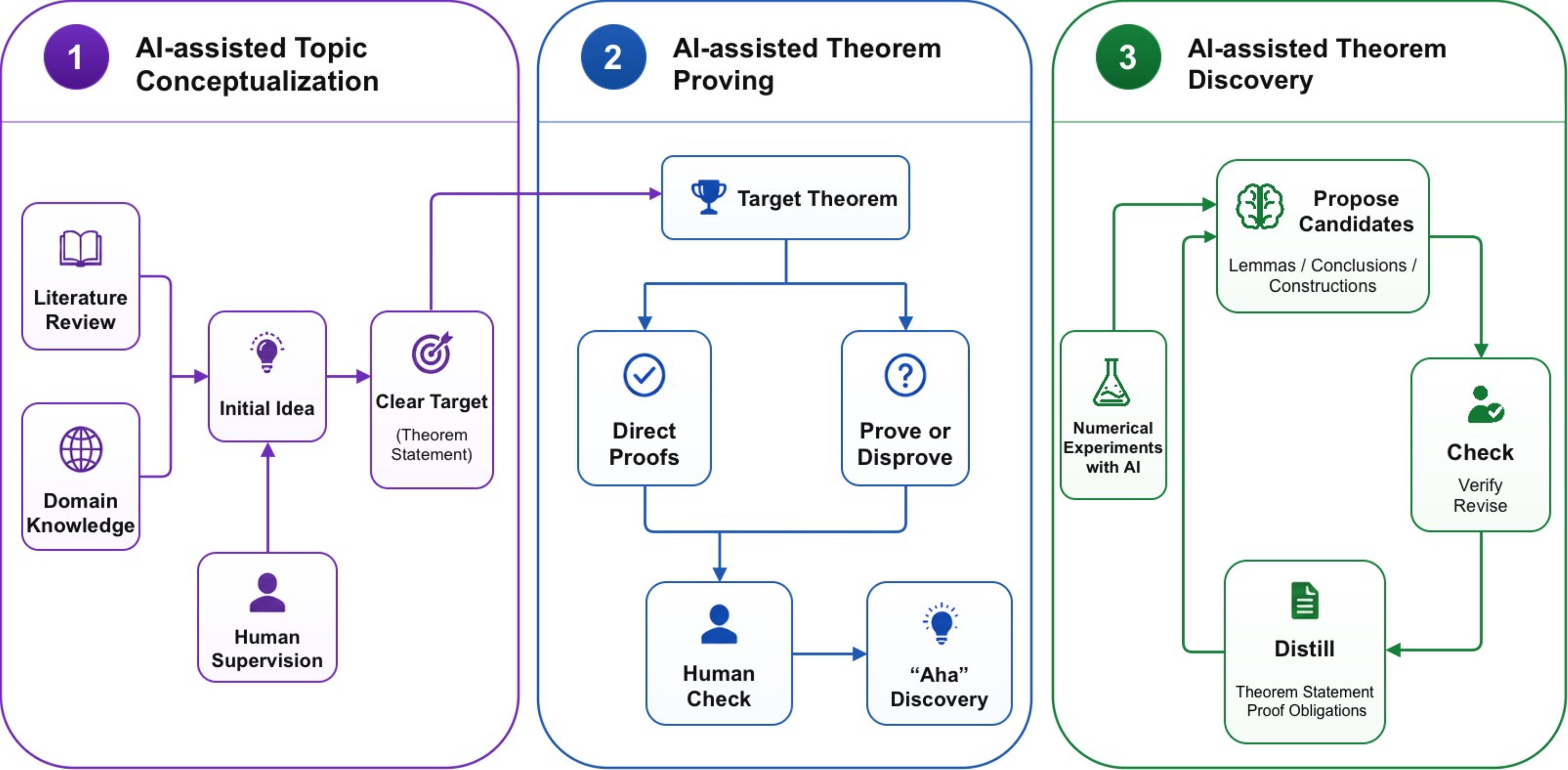}
    \caption{Human-AI interactive theorem proving pipeline. Model proposals, human checks, and distilled theorem statements are organized across the three workflow stages.}
    \label{fig: pipline}
\end{figure}

These observations motivate our interactive workflow, which inserts human oversight at intermediate stages and uses computationally reproducible checks and explicit proof obligations to support the conclusions. We document how this workflow converts model interactions into candidate definitions, lemmas, proof drafts, and failure records that experts can check across three case-study threads. Our human-AI workflow centers on three modes of human--AI interaction, namely conceptualizing research topics, theorem proving for fixed targets, and theorem discovery while the conclusion remains open. As the main case study, we study Grover's quantum search algorithm through Riemannian optimization. We explicitly present the prompts that instantiate the workflow in this setting. The human-AI interactive proving pipeline is shown in Figure \ref{fig: pipline}. Our main contributions are listed as follows.
\begin{enumerate}
  \item \textbf{Topic conceptualization with AI.} We design an AI-assisted routine for early stage problem formulation. Starting from preliminary human ideas about research topics, the LLM proposes candidate objects, assumptions, and questions with brief notes on novelty and feasibility. Researchers, guided by both their own expertise and the model's feedback, test these candidates on minimal examples. The surviving ideas are turned into precise statements and an initial paper outline.

  \item \textbf{Goal-guided proving.} For clearly specified proof targets, researchers communicate their preliminary proof ideas and directions to the model via carefully designed prompts. The LLM suggests reductions, proof sketches, and possible counterexamples along these directions. After human checks, we extract clear theorem statements, proofs, and auxiliary lemmas, and consolidate them into rigorous, human-readable theorems.

  \item \textbf{Open-ended theorem discovery.}  When target conclusions remain open, we use the LLM to propose candidate statements with lightweight evidence, such as numerical experiments and mathematical reasoning. Human experts filter and refine these candidates into theorem templates and reusable proof schemas for related problems. When suitable constructions are unknown, a loop of property-constrained generation and testing searches for simple candidates that satisfy the desired conditions and admit clean proofs.

  \item \textbf{Documented case studies in mathematical optimization.} We use two related Grover-compatible threads and a third composite convex-optimization case to record how the workflow operated on concrete proving and discovery tasks. The model proposes proof routes, structural observations, or candidate updates. The human experts test these suggestions by calculation, compatibility with the surrounding theory, and proof reconstruction.
\end{enumerate}

Table~\ref{tab:statement-provenance} gives statement-level provenance for the mathematical results used in the case-study sections. Results marked as reproduced or adapted originate in the companion Grover-compatible optimization article \cite{lai2025grover} and are included here as mathematical material for documenting the corresponding human--AI research process. Statements marked as new to this manuscript are introduced in the present workflow exposition.  The workflow's generalizability beyond the documented case studies remains untested.

We organize this paper according to different modes of AI-assisted mathematical practice. We briefly review elements of mathematical research in Section \ref{sec:pre}. Topic conceptualization with AI support is discussed in Section \ref{sec: topic}. We present AI-assisted theorem proving for fixed targets and document the associated records that human experts can check in Section \ref{sec: proof}. AI-assisted theorem discovery, including proposal, numerical test, and proof, together with transfer to similar settings, is explored in Section \ref{sec: discover}. We clarify the respective roles of human experts and AI systems in Section \ref{sec:roles}. Finally, the paper is concluded with Section \ref{sec: conclusion}. Appendix~A supplements the main case study with a related multi-oracle case, representative failed and corrected routes, and the Grover-compatible prompt records. Appendix~B contains another separate composite-optimization case study.

\section{Preliminaries for mathematical research}
\label{sec:pre}

In practice, mathematicians often view research as passing through five connected phases, namely {topic conceptualization}, {conjecturing}, {validation}, {proof development}, and {writing}. Each phase has its own goals, typical outputs, and methods. We briefly review the role of each phase and indicate how new advances can be categorized.

We use the following notation consistently in the case study. The unitary group is denoted by $\mathrm{U}(N)$. A ket $|\psi_k\rangle$ denotes a unit vector, whereas $\psi_k:=|\psi_k\rangle\langle\psi_k|$ denotes the associated rank-one projector. Success probabilities are written as $q_k=\operatorname{Tr}(H\psi_k)$, equivalently $q_k=\langle\psi_k|H|\psi_k\rangle$ when the ket notation is used explicitly. Gradients are Riemannian gradients on $\mathrm{U}(N)$ unless the text explicitly refers to the skew-Hermitian representative of the gradient.

\textbf{Topic conceptualization.}
The first task is topic conceptualization. Good topics balance depth, tractability, and potential for broader impact \cite{tao2007good}. The conceptualization process specifies the scope, modeling decisions, and targets for correctness and clarity. Researchers begin by specifying the object, the assumptions, and the claim. The object identifies the mathematical structure under study, such as a class of functions, graphs, or manifolds. The assumptions collect the standing hypotheses, for example, the regularity conditions, dimensional regimes, or model constraints. {The claim formulates the intended conclusion in precise terms, such as an error bound, a convergence rate, or a stability guarantee.} In practice, this stage includes surveying prior work, reviewing both methods and known limitations, and choosing the working setting, such as toy models for quick feedback or general frameworks for the final theorem.

\textbf{Conjecturing.}
Once the main topic is relatively fixed, human experts study the related literature and learn the relevant concepts. They ask what existing results already cover, where the assumptions are too strong, which regimes or examples are missing, and how known methods might be modified. They also look for ideas from neighboring areas that could be imported into the current setting. From this perspective, conjectures often come from extending an existing theorem to a new regime, weakening assumptions while trying to keep the same type of conclusion, or combining mechanisms that have been used separately \cite{polya1957solve}.
AI can generate many candidate conjectures and variations, especially when broad background knowledge or cross-domain analogies are useful. The central difficulty is to select candidates that are mathematically important, feasible, and worth developing. This selection still depends on expert judgment.

\textbf{Innovation types in conjecturing.}
We now distinguish several common innovation routes at the conjecturing stage and, for each route, explain how conjectures arise and provide representative examples.

\begin{enumerate}
    \item \textbf{Fix bug.} Make existing methods verifiable or stable under realistic assumptions. Typical examples include robustness to nonsmooth terms and convergence proofs under perturbations, such as proximal steps and operator splitting for nonsmooth objectives.

    \item \textbf{Weaken assumptions.} Replace strong regularity or noise free models with weaker, more realistic conditions while preserving guarantees. For instance, establish convergence or rates under weak convexity, stochastic perturbations, or inexact oracles.

    \item \textbf{Combination fusion.} Merge complementary mechanisms to improve performance. For example, combine acceleration with variance reduction to obtain provable improvements over either component alone.

    \item \textbf{Cross domain lifts.} Recast the problem in a different domain, then use the new viewpoint to propose related questions. Typical cases include passing from the discrete to the continuous setting, and from a Euclidean setting to a manifold one. A standard example outside the present case study is to model an accelerated optimization method by a continuous-time ODE, and then use Lyapunov or energy functions from ODE stability theory to prove convergence properties of the corresponding discrete method.

    \item \textbf{Unifying viewpoints.} Place several related methods in a single framework to explain their similarities and differences. For example, viewing first-order methods through a parameterized ODE can unify different acceleration schemes, which can then be analyzed within a Lyapunov framework.

    \item \textbf{Boundary and negative results.} Identify the limits of what is achievable under a given set of assumptions. Common goals include proving lower bounds and constructing sharp counterexamples that expose boundary cases.
\end{enumerate}

\textbf{Validation.}
Before investing effort in proofs, human experts test candidate conjectures in straightforward ways. They may compare them with known results and basic intuition, and examine their behavior on both simple examples and complicated cases. Feedback from these checks is then used to revise the statement, adjust the assumptions, and decide which directions are worth pursuing.

In practice, human experts combine careful manual calculations with small numerical experiments. Examples that cover boundary cases and important special situations may also be explored. These checks can be implemented with lightweight tools such as Python with the NumPy and SciPy libraries, or MATLAB, depending on the surrounding code base. In the present work, they serve as diagnostic evidence that guides proof attempts. Among the five phases, validation is one of the most difficult to delegate to AI. A model can propose tests, generate executable calculations, or point to apparent patterns, but human experts must execute and assess the checks, determine whether hidden assumptions remain, and decide whether the statement is ready for proof. It is often helpful to keep a record of each conjecture, which contains the tests it has passed or failed, and the exact assumptions under which it seems to hold. The goal is to discard weak ideas early, focus on a clear formulation, and identify proof directions.

\textbf{Proof development.}
When the statement is stable, human experts try to figure out a proof plan.
Several basic proof approaches are useful in many parts of mathematics.
First, human experts use plain step-by-step reasoning. They write down the definitions, expand them carefully, and follow the logical implications until the claim is reached.
Second, they adapt standard methods from previous work or neighboring areas. Typical transfers take techniques developed in other domains and apply them in the current setting. Decomposing the current problem into simpler subproblems can reveal structures that also appear in related work.
Third, they look for innovative structure or ideas that can guide the proof. This may involve defining a quantity with carefully chosen properties. At this stage, new proof techniques may also be developed to handle more sophisticated problems.

\textbf{Writing.}
In the final stage, human experts organize the statements, standardize the notation, and polish the proofs. Results are arranged into a coherent hierarchy of definitions, lemmas, and theorems, and symbols are introduced explicitly and used consistently. Proofs are streamlined where possible, with the main idea stated before technical details. The final writeup gathers the work from conjecturing, testing, and proving into a presentation that is clear, checkable, and easy to build on. {In the case study, the writing-phase outputs are the human-curated theorem hierarchy, standardized notation, polished proofs, and final manuscript. The main case-study analysis centers on AI-assisted theorem proving and theorem discovery.}

\section{Conceptualizing research topics with AI} \label{sec: topic}

\subsection{Early stage exploration with domain sketches and candidate questions}\label{sec: early-stage}
Early exploration moves from broad themes to specific mathematical targets. We begin by using AI to map the structure of the domain. This includes identifying the main objects under study, including core algorithms, key operators, and typical function classes. We also identify the assumptions that usually govern behavior, namely properties of functions, or curvature conditions, and the standard objectives like rates, bounds, stability, and existence results. By aligning terminology across different areas and highlighting analogies, AI helps construct a coherent picture of the field and locate promising candidate problems. A typical example is transferring properties established in Euclidean space to their counterparts on manifolds.

With this guidance, we encourage AI to propose concrete research questions along principled axes of variation. These include weakening or strengthening hypotheses, transferring known results to new regimes, seeking converses, or isolating thresholds. Each proposal is paired with a precise claim sketch with minimal hypotheses, the nearest classical results with their gaps, and a plausible method class. To avoid vacuous or ill-posed targets, we run quick feasibility checks. AI is asked to generate instances, and to perform scaling and invariance sanity checks. These steps reveal hidden necessary conditions or point to underlying patterns. Proposals that survive are ranked by novelty and tractability, with preference given to those promising a unifying principle. The output of this stage is a compact domain sketch and a curated set of viable questions ready to be sharpened.

\textbf{Example.} Building on the perspective of \cite{suzuki2025grover}, which analyzes classical quantum algorithms through imaginary time evolution, we begin with an initial idea of combining quantum algorithms with manifold optimization. We phrase this idea as a concrete prompt, given in Appendix \ref{appendix: early}. The model returns a concise conclusion that lists candidate directions and prerequisite concepts on both sides, such as algorithmic primitives, target objectives, and possible correspondences between them, together with preliminary feasibility explanations. We treat this response as a rough map. It guides our literature reading and internal discussion, and helps us prune ideas that are ill-posed or unlikely to admit clean analysis. After several rounds of interaction with language models, our main idea converges on two anchors, Grover's search and Riemannian gradient methods.

\subsection{Concrete idea phase for sharpened claims and validation planning} \label{sec: concrete-idea}

Once the topic and basic directions are in place, the next step is to turn promising ideas into precise and checkable statements. Human experts usually begin by fixing notation and writing down assumptions as explicit structural properties of the objects under study. This separates the object of analysis from background motivation. At the same time, AI can be asked to propose decompositions, identify plausible proof routes, and point out where known techniques are likely to apply. The goal is to expose the core difficulty of the problem and to identify a small number of central conjectures and theorems that capture it.

With candidate statements in hand, we organize the reasoning into a hierarchy of proof tasks. This hierarchy includes main theorems, key reductions, comparison steps, and delicate lemmas where constants, parameter ranges, or limiting regimes matter most. For each task, AI can list possible proof strategies and propose simple numerical tests. These tests help diagnose which parts are routine and which are genuine bottlenecks. On this basis, we design a validation plan that separates target claims from structurally suggested conjectures.

Finally, we start to shape the final paper around this structure. Definitions and notation are organized in the form most convenient for the subsequent arguments. Intermediate claims are grouped into coherent subsections. Prospective figures and tables are tied directly to specific arguments or comparisons. By the end of this phase, a vague direction has been converted into a disciplined roadmap with clear scope, proof avenues, and validation checks.

\textbf{Example.} Continuing the example from the previous subsection, suppose we have fixed the topic at the intersection of Riemannian optimization and Grover's algorithm. The guiding theme is to interpret Grover's procedure from a manifold optimization viewpoint. With this topic in place, we ask the model for a theoretical framing. It produces a structured note that lists the relevant objects on both sides. These include the manifold of quantum states and the standard Riemannian gradient updates. It also recalls the basic properties of Grover iterations, namely their symmetry and complexity guarantees. It suggests natural submanifolds on which to restrict the dynamics and identifies candidate correspondences between gradient dynamics and Grover iterations. The prompts used in this example are given in Appendix \ref{appendix: sharpened claims}.

We then request a more explicit manifold-optimization treatment. The model proposes viewing the state evolution as gradient-type updates on a suitable manifold of quantum states and outlines checks that distinguish genuine geometric structure from artifacts of a particular parameterization. It suggests candidate statements that aim to recover the classical complexity of Grover's algorithm within this manifold-gradient framework. Human authors then refine these suggestions into precise assumptions, theorem statements, and a short list of claims that still need proof.

\section{AI-assisted theorem proving} \label{sec: proof}

This section considers three fixed-target proving scenarios. Human experts specify the goals, choose and verify proof routes, and integrate accepted arguments, while the model proposes local lemmas, alternative approaches, and computational checks.
Several substantive Grover-compatible mathematical results used in the following case-study sections, including the retraction, invariant-structure results, and convergence analysis, originate in or are adapted from the companion mathematical article \cite{lai2025grover}. Their role here is to provide mathematical material for documenting the corresponding human--AI research process.

\begin{table}[htbp]
\centering
\scriptsize
\renewcommand{\arraystretch}{1.12}
\setlength{\tabcolsep}{3pt}
\manualtablecaption{tab:statement-provenance}{Statement-level provenance. \emph{Reproduced}: the same statement appears in \cite{lai2025grover}. \emph{Adapted}: based on material in \cite{lai2025grover} but reorganized, combined for the present exposition. \emph{New}: introduced here, with the label indicating provenance.}
\begin{tabularx}{0.98\linewidth}{>{\raggedright\arraybackslash}p{0.24\linewidth}>{\raggedright\arraybackslash}p{0.14\linewidth}X}
\hline
Statement & Status & Provenance and role \\
\hline
Theorem~\ref{thm: optimality} & Reproduced & The corresponding global-optimality statement and gradient characterization appear as Theorem~2 in \cite{lai2025grover}. Notation is harmonized here. \\
Proposition~\ref{prop: uncertain} & New to this manuscript & This commutator characterization is introduced as a prove-or-disprove example. It is a standard matrix/Lie-algebra fact and is not claimed as a new mathematical research contribution. \\
Theorem~\ref{thm: aha} & Adapted & Based on the classical-simulability result, Theorem~13 in \cite{lai2025grover}, and reorganized around the scalar process $(x_k,y_k,q_k)$. \\
Proposition~\ref{thm:exp-update} & Adapted & The exact exponential-step statement combines the commutator-norm identity and invariant two-dimensional dynamics in \cite{lai2025grover} with the one-step formulation used here. \\
Theorem~\ref{thm:inv-2d} & Reproduced & The invariant-$2D$-subspace and gradient-subspace statements appear as Theorem~6 in \cite{lai2025grover}. \\
Theorem~\ref{thm: 5-factor} & Reproduced & The 5-factor Grover-compatible retraction and its parameterization appear in Proposition~10 of \cite{lai2025grover}. \\
Theorem~\ref{thm-bese-complexity} & Reproduced & The baseline smoothness-based Riemannian-gradient complexity statement appears as Theorem~17 in \cite{lai2025grover}. \\
Theorem~\ref{thm-5factor-bounds} & Reproduced & The tight first- and second-order bounds and constants appear as Lemma~15 in \cite{lai2025grover}. \\
Theorem~\ref{thm-best-complexity} & Adapted & The PL-based complexity statement is based on Proposition~18 and Theorem~19 in \cite{lai2025grover}. \\
\hline
\end{tabularx}
\end{table}

\subsection{Direct proofs when the statements are trusted}\label{sec:direct proofs}
The easiest case begins with a statement human experts believe to be correct and a concrete proof plan. The simplest way to use an AI prover in this setting is to ask it directly for a proof of the target statement, then query the model for standard reductions and a stepwise derivation. Because the model may skip justifications, misapply lemmas, or mishandle quantifiers and constants, human experts use targeted prompts. For example, they explicitly state assumptions and variable domains, require line-by-line reasoning, and ask for citations, adversarial checks, and quick numerical spot tests. Human experts then verify every claim produced by the AI, checking edge cases and reproducing key calculations.

This setup typically arises when human experts already have a deep understanding of what needs to be proved and only some local details remain to be filled in. After reviewing the draft, they identify gaps, remove redundancy, and reorganize the argument so that the key steps are transparent. In this setting, AI is useful primarily as a generator of local proof variants and checks. Human experts provide validation, judgment, and final integration into a clear and fully rigorous proof.

\textbf{Example.}
We illustrate our human-AI workflow with the following claim, which ties stationarity to global optimality on the unitary manifold. Let $H = H^\dagger = H^2$ be an orthogonal projector on a Hilbert space. $U \in \mathrm{U}(N)$ denotes the quantum circuit we want to design. Our cost function is $f (U)=\Tr(HU\psi_0U^\dagger)$, where $\left|\psi_0\right\rangle$ is the initial state. Intuitively, the skew-Hermitian component of the Riemannian gradient $\widetilde{\operatorname{grad}} f(U) = [H,\psi_U]$ shows that stationarity is equivalent to the output state $\psi_U$ commuting with $H$. In this case $\psi_U$ is an eigenstate of $H$, and since the spectrum of $H$ is $\{0,1\}$, the expectation $f(U)=\operatorname{Tr}(H\psi_U)$ is forced to take an extremal value $0$ or $1$. This intuition gives the following theorem on global optimality.

\begin{theorem}[Global optimality]\label{thm: optimality}
Let $f(U) = \operatorname{Tr}(H U \psi_0 U^\dagger)$ on the unitary group
$\mathrm{U}(N)$, and write $\psi_U := U \psi_0 U^\dagger$ for the output
state of the circuit $U$. The skew-Hermitian part of the Riemannian gradient
of $f$ on $\mathrm{U}(N)$ is
\[
  \widetilde{\operatorname{grad}} f(U) = [H,\psi_U].
\]
Moreover, $\widetilde{\operatorname{grad}} f(U) = 0$ if and only if $U$ is a
global optimizer of $f$ over $\mathrm{U}(N)$. In that case
$f(U) \in \{0,1\}$, corresponding respectively to the global minimum $0$
and global maximum $1$.
\end{theorem}

To turn the intuition into a rigorous argument, we use a structured prompt that asks the model to restate assumptions, notation, and domains, give a stepwise derivation, and justify nontrivial inferences. Reference verification is a key methodological safeguard: the model may invoke standard named results by their usual names, while human experts verify any bibliographic support before it enters the proof. Human experts also verify the derivation line by line, reproduce key calculations, check boundary cases, determine citation status, and streamline the final proof. The prompts used in this example are listed in Appendix~\ref{appendix:direct proofs}.

For Theorem~\ref{thm: optimality}, one concrete adversarial check uses the two-dimensional projector \(H=\operatorname{diag}(1,0)\). For the pure states \(\psi=e_1e_1^\dagger\) and \(\psi=e_2e_2^\dagger\), one has \([H,\psi]=0\) and \(\operatorname{Tr}(H\psi)=1\) or \(0\), respectively, so the stationary cases are exactly extremal in this test. For the nonextremal pure state \(v=(e_1+e_2)/\sqrt2\) and \(\psi=vv^\dagger\), one has \(\operatorname{Tr}(H\psi)=1/2\) and \([H,\psi]\ne0\). This confirms that the stationary classification selects the two extremal states in this test.

\subsection{Clear target, uncertain truth}\label{sec: clear target}
In many cases, the truth of a proposed theorem remains uncertain, so we adopt a systematic prove-or-disprove protocol. We delegate to AI along two coordinated branches. One explores plausible avenues toward a proof, and the other actively searches for failure. Both outcomes are informative. A successful argument consolidates the scope of the claim, while a counterexample clarifies its limits and may suggest a sharper statement.

On the proof side, the AI proposes feasible directions, candidate lemmas, decomposition strategies, and intermediate claims, around which the human expert can organize a coherent argument. On the disproof side, the AI generates tentative constructions, stress tests, and diagnostic patterns that focus the search for counterexamples on promising regions of the space. By jointly assessing the evidence produced by both branches, the human expert avoids one-sided reasoning and moves more systematically from an initial conjecture to a stable, well-justified conclusion.

\textbf{Example.} We illustrate the prove or disprove workflow on a basic structural question from matrix theory related to the retraction on the Riemannian manifold. Consider the following proposition about commutators of Hermitian matrices and the Lie algebra $\mathfrak{su}(n)$. It is straightforward to check that any commutator $[A,B]$ with $A^\dagger = A$ and $B^\dagger = B$ is skew-Hermitian. The converse direction is less obvious. Every skew-Hermitian matrix with zero trace must be represented in this way. This makes the statement a natural candidate for a two branch search. The proof branch looks for a constructive representation of arbitrary elements of $\mathfrak{su}(n)$ as commutators, while the disproof branch asks the model to search for counterexamples or obstructions.

\begin{prop}\label{prop: uncertain}
The set of commutators of Hermitian matrices
$$
    \mathcal{C}=\{[A, B]: \, A^\dagger=A, \; B^\dagger=B\}
$$
equals
$
    \{X\in M_n (\mathbb{C}): \, X^\dagger=-X, \; \mathrm{tr}\, X=0\}=\mathfrak{su} (n).
$
\end{prop}

To analyze this proposition, we use a prove-or-disprove prompt with two branches. The proof branch asks for a construction of arbitrary elements of $\mathfrak{su}(n)$ as commutators $[A,B]$. The disproof branch searches for obstructions, counterexamples, and invariant-based failures. Each branch must give explicit reasoning and use only standard results. The output is a verdict together with adversarial checks that human experts inspect before accepting the conclusion. Concrete prompts are given in Appendix \ref{appendix: uncertain}.

\subsection{Rich proofs and the ``Aha'' moment}\label{sec: rich proofs}

A model response may contain candidate invariants, reductions, or alternative formulations beyond the requested proof. We document their provenance by using prompt records to identify the initial observation and the manuscript to record the human work of selecting, formalizing, verifying, and integrating it.

\textbf{Example.}
In the synthesis example, the model's reply included a secondary observation about the update dynamics. The AI-generated seed is identifiable in Appendix~\ref{appendix: rich}, where the model pointed to confinement in a two-dimensional working subspace. The human contribution lies in reworking that seed into a mathematical result. Human authors recognized that the high-dimensional unitary evolution could be tracked through a small number of scalar variables once the dynamics was restricted to the Grover plane. They then formulated a theorem that records the associated classically computable scalar process and resolves the matrix product into a low-dimensional update.

\begin{theorem}\label{thm: aha}
Let $H$ be a projector and $\psi_0$ a rank-one projector. Assume $0<\operatorname{Tr}(H\psi_0)<1$. Fix the five-factor gate-product curve $V(t;x,y)$ of Theorem~\ref{thm: 5-factor}, with parameters prescribed by \eqref{eq: 5 parameter} and $V(t;0,0)=I$. Define $U_0=I$, $\psi_k=U_k\psi_0U_k^\dagger$, and
\[
\begin{gathered}
{}[H,\psi_k]=x_kX_0+y_kY_0,\\
V_k=V(t_k;x_k,y_k),\qquad U_{k+1}=V_kU_k,
\end{gathered}
\]
where $X_0=[H,\psi_0]$ and $Y_0=i[H,X_0]$.
Then there exists an explicit, classically computable process
\begin{equation}\label{eq-F}
    F_{t_k}(x_k, y_k, q_k)=(x_{k+1}, y_{k+1}, q_{k+1}),
\end{equation}
where $(x_k,y_k)\in\mathbb{R}^2$ are the real coordinates of the commutator $[H,\psi_k]$ in the fixed basis $(X_0,Y_0)$ of the two-dimensional gradient plane, $q_k=\operatorname{Tr}(H\psi_k)\in[0,1]$ is the scalar success probability, and $t_k\in\mathbb{R}_{\geq0}$ is the stepsize. For this fixed product and parameter rule, $F$ is a three-dimensional real scalar update on $\mathbb{R}^2\times[0,1]$, parametrized by $t_k$ and obtained by restricting the full $N\times N$ matrix dynamics to the Grover plane. This scalar process keeps all iterates within $\mathrm{U}(N)$, preserves the Grover plane $\mathcal{S}$, and admits closed-form scalar recursions.
\end{theorem}

\section{AI-assisted theorem discovery} \label{sec: discover}

Theorem discovery is often harder than fixed-target proving because the final statement remains open during the search. AI can help by proposing candidate statements, patterns, invariants, normal forms, and boundary tests. Human experts still fix the assumptions, judge the evidence, and decide which byproducts are promoted to named results.

\subsection{From assumptions to conclusions through AI-guided discovery}\label{sec:assumption}

\subsubsection{AI-guided proposal of candidate conclusions}
\label{sec: AI-guided proposal}

Human experts begin with a fixed setting, assumptions, class of objects, and admissible ambient space, while the precise {conclusion} remains open. In this situation, they prompt the AI to {propose} candidate consequences of the assumptions, such as plausible properties, equivalent formulations, reductions, and statements about extreme or boundary regimes. To encourage diversity, they ask for multiple variants with short explanations and, when appropriate, suggested counterexamples. Since the model can make mistakes or overstate what is true, each proposal must provide light supporting evidence, such as tests on simple and edge cases, quick numerical checks, and references to comparable results. The returned suggestions are then filtered by explicit criteria, and for each candidate we record whether it passes or fails and why.

Human experts then {analyze and select}. They test validity more thoroughly, evaluate whether a candidate improves interpretability or tightens bounds, and keep the statements that clarify mechanisms or simplify later analysis. From these surviving candidates, they distill a clean, testable theorem shape, including assumptions and theorems. AI is used to expand it into a coherent, step-by-step argument. Discarded claims are recorded as negative evidence and inform the next round of prompts.

\textbf{Example.}
We illustrate the discovery workflow on the Grover-type iteration.
Our goal is to explore the properties of a simple projector-driven dynamics underlying the Grover search operator. We use the AI interaction to surface candidate structural statements. At a high level, we study the evolution of pure states under repeated updates driven by a fixed projector and ask for invariants, norm identities, and scalar recursions that describe this process.

First, we consider the general exponential update. Concretely, fix a projector $H=H^\dagger=H^2$ and consider pure states $\psi_k=|\psi_k\rangle\langle\psi_k|$. Define $q_k=\operatorname{Tr}(H\psi_k)\in[0,1]$ and $\gamma_k=\sqrt{q_k(1-q_k)}$. We study the following {exact exponential step}
\begin{align}\label{eq: exact step via exponential}
      V_k=\exp\bigl(t_k[H,\psi_k]\bigr),\qquad U_{k+1}=V_k U_k,\qquad \psi_{k+1}=U_{k+1}\psi_0 U_{k+1}^\dagger.
\end{align}
At this stage, the structurally decisive properties of this update are still being identified, so the final theorem is left open. We ask the model to surface plausible consequences of the dynamics, such as invariance properties, norm identities, scalar recursions, or one step convergence conditions. These suggestions are then checked for internal consistency and usefulness for the later analysis, retaining only statements that clarify the mechanism or simplify the argument. This leads to the following precise statement.

\begin{prop}[Exact step via exponential (``Exp update'')]\label{thm:exp-update}
Let $H=H^\dagger=H^2$ and define $q_k=\operatorname{Tr}(H\psi_k)$ and $\gamma_k=\sqrt{q_k(1-q_k)}$. The following properties hold for \eqref{eq: exact step via exponential}.
\begin{itemize}
    \item[(i)] The gradient norm identity $\|[H,\psi_k]\|_F=\sqrt{2}\,\gamma_k$.
    \item[(ii)] The state $|\psi_k\rangle$ remains in the real two dimensional Grover plane spanned by $\{|\psi_0\rangle,\, H|\psi_0\rangle\}$.
    \item[(iii)] The quantity $q_{k+1}$ has a closed form recursion obtained by viewing the update as a rotation on this plane.
    \item[(iv)] If $0<q_0<1$, there exists a positive one-shot step
$
t_0^*=\frac{\arccos\sqrt{q_0}}{\gamma_0}
$
such that $e^{t_0^*[H,\psi_0]}|\psi_0\rangle=|\psi^*\rangle$ and $\max f(U)=1$ are attained in a single update. At $q_0=0$ or $q_0=1$, $[H,\psi_0]=0$, so every finite exponential step leaves the state unchanged. The case $q_0=1$ is already optimal, whereas $q_0=0$ cannot reach $f=1$ by this update.
\end{itemize}
\end{prop}

We used the assumption-driven prompt in Appendix~\ref{appendix: proposal} to generate candidate structural properties. Human authors retained only statements that passed the endpoint checks $q_k=0,1$, low-dimensional comparisons between the full matrix update and scalar recursion, and tests outside the stated product-generated Grover plane. These surviving claims form Proposition~\ref{thm:exp-update}.

\subsubsection{Human abstraction and cross-setting transfer}
\label{sec: human abstraction}

After one iteration of the discovery loop, human experts {extract patterns} such as recurring steps, reusable lemmas, and canonical normalizations. These ingredients are summarized into a portable {proof schema} that records what to assume, what to try first, which invariants to test, and which failure modes to monitor. The schema is then {transferred} to nearby settings, such as related function spaces, modified boundary conditions, or perturbed parameter ranges. Human experts prompt the AI to adapt earlier lemmas, inequalities, and constants to generate variants on similar cases.

In summary, the working loop for exploring conclusions of fixed assumptions is as

\begin{center}
$\text{propose} \;\rightarrow\; \text{check} \;\rightarrow\; \text{distill} \;\rightarrow\; \text{prove} \;\rightarrow\; \text{transfer},$
\end{center}
with the AI broadening the search and providing evidence-backed suggestions, and human experts supplying direction, skepticism, selection, and cross-setting reuse. The outcome combines a theorem for the initial setting with a transferable pattern that can be tested in related regimes.

\begin{wrapfigure}[18]{r}{0.36\textwidth}
    \centering
    \setlength{\abovecaptionskip}{4pt}
    \includegraphics[width=\linewidth]{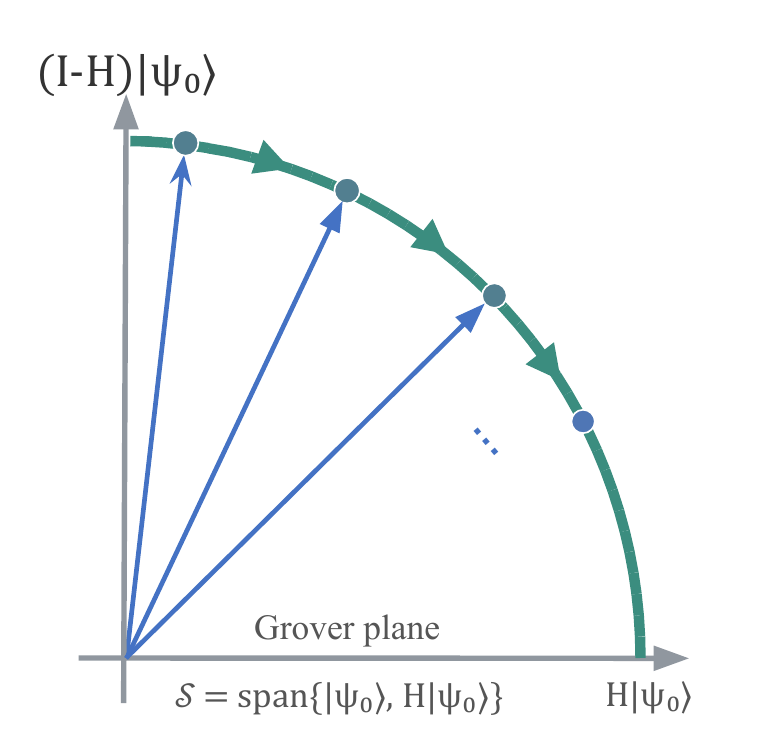}
    \caption{Invariant Grover plane shown using the orthogonal components \(H|\psi_0\rangle\) and \((I-H)|\psi_0\rangle\). Product updates preserve the same subspace \(\mathcal S\).}
    \label{fig:grover-plane}
\end{wrapfigure}

\textbf{Example.}
We treat the projector-driven instance in Section \ref{sec: AI-guided proposal} as an exemplar for transfer. From the prior round, the AI returned several properties that survived lightweight checks by human experts, including an invariant working subspace, a stable two-direction gradient plane, a scalar progress statistic with a simple recursion, and a local surrogate update that preserves structure within controlled error. We abstract these survivors into a portable schema and use them to organize proofs in similar settings. Concretely, we keep the invariance and closure statements as the core of the schema, and ask the model to instantiate them under new assumptions while providing the same evidence pack, consisting of toy and edge-case numerics, minimal counterexamples, and references to comparable mechanisms. The central lemma we retain is as follows.

\begin{theorem}[Invariant $2$D subspace lemma]\label{thm:inv-2d}
Let $H$ be a projector, $\psi_0$ a rank-one projector, and define iterates as follows. Let $V_k$ be any finite product of factors of the form $e^{i\theta H}$ and $e^{i\theta \psi_0}$, set $U_{k+1}:=V_k U_k$ with $U_0=I$, and let $\psi_k:=U_k\psi_0 U_k^\dagger$. Let
\[
    q_k :=\operatorname{Tr}(H\psi_k)\in[0, 1],
\]
and define
\(
    X_0 :=[H, \psi_0], Y_0 :=i[H, X_0].
\)
Then the following properties hold
\begin{itemize}
  \item The state $|\psi_k\rangle$ remains in the fixed two dimensional subspace (the Grover plane) \\
  \(
    \mathcal{S} :=\operatorname{span}_\mathbb{C}\{|\psi_0\rangle, \, H|\psi_0\rangle\}
    =\operatorname{span}_\mathbb{C}\{(I-H)|\psi_0\rangle, \, H|\psi_0\rangle\}
  \) for all $k$.
  \item The skew-Hermitian representation of each gradient satisfies
  $\widetilde{\operatorname{grad}} f(U_k)=[H,\psi_k]\in\mathcal W$, where $\mathcal W:=\operatorname{span}_{\mathbb R}\{X_0,Y_0\}$.
  Equivalently, the tangent gradient is
  $\operatorname{grad} f(U_k)=[H,\psi_k]U_k$, which belongs to $\mathcal WU_k$ for all $k$.
\end{itemize}
\end{theorem}

Figure~\ref{fig:grover-plane} illustrates this invariant plane using the orthogonal decomposition \( |\psi_0\rangle=(I-H)|\psi_0\rangle+H|\psi_0\rangle\).

The prompt is given in Appendix \ref{appendix: human abstraction}. For cross-setting transfer, Proposition \ref{thm:exp-update} is used as a base lemma. The model is asked to identify an invariant working subspace, a scalar progress statistic, and a local surrogate update. Possible transfer settings include variational quantum eigensolver dynamics, QAOA-style product ansatzes, and amplitude-amplification variants with different projectors. We treat these as proposed directions for future validation. The proof of Theorem~\ref{thm:inv-2d} was then written and checked under the same human-verified workflow.

\subsection{Property constrained synthesis} \label{sec: synthesis}
Human experts specify the required properties, admissible space, and edge cases. The model proposes parameterized candidates together with executable tests. Human authors run the tests locally, compare candidates by correctness and simplicity, and send only surviving forms to symbolic proof. Failed candidates lead to tighter assumptions or a revised search.

\textbf{Example.}
We treat the design of the retraction in our update rule as a \emph{property-constrained search} guided by the AI. The target properties are fixed in advance.
\begin{itemize}
  \item[(P1)] $V(t;x,y)$ is a product of factors of the form $e^{i\theta H}$ or $e^{i\theta\psi_0}$.
  \item[(P2)] $V(0;x,y)=I$.
  \item[(P3)] for any target tangent $Z\in\mathcal W:=\mathrm{span}_\mathbb{R}\{X_0,Y_0\}$ with $Z=xX_0+yY_0$ (where $X_0:=[H,\psi_0]$, $Y_0:=i[H,X_0]$), the initial velocity satisfies $V'(0;x,y)=Z$.
\end{itemize}

We ask the model to propose product forms and parameter couplings of different lengths and to supply executable checks for each candidate on a seed set $\mathcal S\subset\mathbb R^2$. The supplied tests check (i) (P1) syntactically, (ii) (P2) exactly, and (iii) (P3) by the finite-difference criterion
\begin{equation}\label{eq: test}
    \left\|\frac{V(h;x,y)-I}{h}-(xX_0+yY_0)\right\|_F\le\varepsilon,
\end{equation}
for a small step size $h$ and each $(x,y)\in\mathcal S$. Human authors execute these tests locally, including the edge cases $x=0$ and $y=0$. Only the shortest product that passes this screening is sent to detailed proof.

The following summary records the candidate screening stage. The search objective was to find a product retraction with as few factors as possible while preserving \textup{(P1)}-\textup{(P3)} and admitting a clean proof. The length-3 and length-4 candidates were rejected by the checks. The length-5 product was the shortest construction that passed numerical screening and was then verified symbolically. Longer length-6 and length-8 block constructions can also be written down, but they add factors while leaving the present proof unchanged. Table~\ref{tab:retraction-candidates} summarizes these candidates, where \(C_{a,b}:=e^{iaH}e^{ib\psi_0}e^{-iaH}\).

\begin{table}[htbp]
\centering
\manualtablecaption{tab:retraction-candidates}{Candidate product retractions in the synthesis loop.}
\small
\renewcommand{\arraystretch}{1.16}
\setlength{\tabcolsep}{4pt}
\begin{tabular}{@{}>{\centering\arraybackslash}p{0.08\linewidth} >{\raggedright\arraybackslash}p{0.50\linewidth} >{\raggedright\arraybackslash}p{0.32\linewidth}@{}}
\hline
Factors & Representative product & Conclusion \\
\hline
\(3\)--\(4\) & Short alternating products in \(H\) and \(\psi_0\). & Rejected by the checks. \\
\(5\) &
\(\begin{aligned}
&e^{ia_1H}e^{ib_1\psi_0}e^{i(a_2-a_1)H}e^{ib_2\psi_0}e^{-ia_2H}U
\end{aligned}\)
& Shortest validated candidate used in the proof. \\
\(6\) & \(C_{\pi/2,c_1}C_{-\pi/2,c_2}C_{0,y}U\). & Valid longer construction with the same proof role. \\
\(8\) &
\(\begin{aligned}
&C_{0,y/2}C_{\pi/2,-x/2} C_{-\pi/2,x/2}C_{-\pi,-y/2}U
\end{aligned}\)
& Valid longer construction with the same proof role. \\
\hline
\end{tabular}
\end{table}

\begin{samepage}
\begin{theorem}\label{thm: 5-factor}
{For $U\in\mathrm U(N)$ and $\eta=ZU\in\mathcal WU$, with $Z=xX_0+yY_0$,} define the following length-5 product retraction
\begin{equation}\label{eq: 5-factor}
    \mathrm{R}^{(5)}_U (\eta): =e^{i a_1 H} e^{i b_1 \psi_0} e^{i\left (a_2-a_1\right) H} e^{i b_2 \psi_0} e^{-i a_2 H} \, U,
\end{equation}
where the real parameters $(a_1,a_2,b_1,b_2)$ are defined as
\begin{align}\label{eq: 5 parameter}
\begin{aligned}
    A&:=\operatorname{atan2} (y, x), \quad R:=\sqrt{x^2+y^2},\quad a_1=A+\frac{\pi}{2},\\ a_2&=A-\frac{\pi}{2}, \quad b_1=-\frac{R}{2}, \quad b_2= \frac{R}{2}.
\end{aligned}
\end{align}
{When $(x,y)=(0,0)$, set $\mathrm R^{(5)}_U(0)=U$ and $V(t;0,0)=I$. For $(x,y)\ne(0,0)$, the gate-product curve}
\begin{equation*}
{V(t;x,y)}={e^{i a_1 H} e^{i t b_1 \psi_0} e^{i\left(a_2-a_1\right) H} e^{i t b_2 \psi_0} e^{-i a_2 H}}, \quad t \geq 0,
\end{equation*}
satisfies (P1)--(P3), with $V(0;x,y)=I$ and $V'(0;x,y)=Z$. The corresponding retraction curve through $U$ is
{
\[
\gamma_U(t;x,y):=\mathrm R^{(5)}_U(t\eta)=V(t;x,y)U,
\qquad \gamma_U(0;x,y)=U,\quad \gamma_U'(0;x,y)=\eta.
\]
}
\end{theorem}
\end{samepage}

The full property-constrained prompt and test specification are given in Appendix~\ref{appendix: property constrained}.
The selected length-5 product is the shortest robust candidate in the recorded search. Human experts replace the model's proof sketch with a symbolic derivation and integrate it into the main argument. The reduction uses that $H$ is a projector, $\psi_0$ is rank one, and the update products are generated only by $H$ and $\psi_0$. Extensions to nonprojector observables, additional noncommuting Hamiltonian terms, or ansatz families outside the Grover plane would require a separate analysis.

\subsection{Human-AI collaborative proof for conjectures}
Proving complex conjectures is sensitive to the domain, geometric setting, and technical assumptions. We organize the collaboration into four stages: (i) fix the objects, ambient space, and assumptions, (ii) identify a literature-guided proof route, (iii) calibrate constants through analysis and locally executed checks, and (iv) refine the stabilized argument into a sharper theorem. The corresponding outputs are a specification checklist, proof skeleton, tested constants and logs, and final theorem.

\subsubsection{Scoping with AI for objects, ambient spaces, and assumptions}
\label{sec: scoping}

Human experts usually start from a rough target theorem in which several concepts and assumptions are still vague. In the first stage, they conduct short rounds of interaction with the AI whose purpose is to make these ingredients precise. In each round, the AI is asked to define the relevant objects, fix the ambient space (for instance, a Euclidean space, a Riemannian manifold, or a function space), specify the metric and the associated notion of gradient, and list all assumptions in a consistent format. The round ends with an explicit checklist of domains of variables, regularity or convexity requirements, boundary cases, and notation. Human experts review this checklist, resolve inconsistencies, and repeat the cycle until the statement is internally consistent and the AI can restate it in the agreed notation without contradictions or hidden assumptions.

\textbf{Example.}
We start from an informal conjecture that gradient descent with the 5-factor retraction on the unitary group should reach a point with gradient norm at most $\varepsilon$ in
\(
    O(\sqrt{N}/\varepsilon^{2})
\)
iterations, for the target function
\(
    f(U) \;=\; \operatorname{Tr} \bigl\{H\, U \psi_0 U^\dagger\bigr\},
\)
defined on the unitary group $\mathrm{U}(N)$ equipped with the canonical Riemannian metric induced by the Frobenius inner product. At this point, the geometric setting is still unresolved. The iteration can be viewed either through an ambient Euclidean lens or through the intrinsic manifold geometry. This choice determines which notion of Lipschitz continuity should control the complexity bound.

After several scoping rounds with the AI, we fix the following setting.
Gradients and norms are understood in this Riemannian sense. Following standard practice in numerical optimization, we expect that some Lipschitz constant of the (Riemannian) gradient will enter the complexity analysis. The output of the first stage is therefore a precise target theorem. It specifies the domain $U \in \mathrm{U}(N)$ with its canonical metric, the objective $f$, the assumption that the Riemannian gradient of $f$ is Lipschitz with constant $L_g$, and the goal of proving an iteration bound of order $O(\sqrt{N}/\varepsilon^ 2)$ for the 5-factor retraction gradient method.

In this phase, the AI expands the informal conjecture into a well-posed statement about $(\mathrm{U}(N),\text{canonical metric})$, the function $f$, and the constant $L_g$ that later appears in the complexity bound. The prompt used for this example is given in Appendix \ref{appendix: scoping}.

\subsubsection{Route confirmation through a literature guided plan}\label{sec: route}

Once the statement is fixed, the second stage searches for a feasible proof route. Human experts read the relevant literature and ask the AI to retrieve related results, known inequalities, and standard proof patterns that match the scoped setting. The goal is to construct a detailed proof skeleton, including a main strategy, a sequence of reductions, and key lemmas and inequalities with clear logical dependencies. The AI is used to suggest plausible decompositions of the main claim, to recall standard complexity bounds under comparable assumptions, and to point to candidate references. Human experts then check compatibility with the scoped statement and record which external results will be instantiated and which conditions remain to be verified.

\textbf{Example.}
For the convergence conjecture above, experience suggests that the iteration complexity of a gradient method is controlled by a Lipschitz constant of the gradient. As a first attempt, we ask the AI to estimate a Euclidean Lipschitz constant of the gradient and to combine it with standard Euclidean gradient descent bounds. The resulting rate has the wrong dependence on $N$ and is closer to $O(N/\varepsilon^2)$ than to the desired $O(\sqrt{N}/\varepsilon^2)$. This shows that the Euclidean Lipschitz notion is mismatched to the target rate in our manifold setting.
For the Grover search instance, let $M:=\operatorname{rank}(H)=\operatorname{Tr}(H)$ be the number of marked basis states among the $N$ basis states, with $1\leq M<N$. The reference state is the uniform superposition $|\psi_0\rangle=N^{-1/2}\sum_{j=1}^{N}|j\rangle$, so $\operatorname{Tr}(H\psi_0)=M/N$.

We then ask the AI to search for complexity results for Riemannian gradient descent with retractions. The prompt is given in Appendix \ref{appendix: route}. After comparing the model's suggestions with the optimization literature and discussion among the authors, we identify the relevant parts of \cite{boumal2019global} as the appropriate basis for the proof. We adopt the standard complexity result for Riemannian gradient descent with a Lipschitz continuous Riemannian gradient. This shifts the focus from a Euclidean Lipschitz constant to the Lipschitz constant $L_g$ of the pullback gradient of $f$ on $(\mathrm{U}(N),\text{canonical metric})$. The desired rate $O(\sqrt{N}/\varepsilon^2)$ will follow if we can prove that the Lipschitz constant of the pullback gradient satisfies
\(
    L_{\mathrm{Rie}}
    \in
    O\left(\sqrt{\frac{N}{M}}\right).
\)
\par
We maximize $f$ by Riemannian gradient ascent:
{
\begin{equation}\label{eq:r4-gradient-iteration}
\begin{aligned}
    \psi_k&:=U_k\psi_0U_k^\dagger,\qquad
    G_k:=[H,\psi_k]=x_kX_0+y_kY_0,\\
    \eta_k&:=G_kU_k=\operatorname{grad}f(U_k),\qquad
    t_k:=L_{\mathrm{Rie}}^{-1},\\
    U_{k+1}&:=\mathrm{R}^{(5)}_{U_k}(t_k\eta_k)=V_kU_k,\\
    V_k&:=e^{ia_{1,k}H}e^{it_kb_{1,k}\psi_0}
    e^{i(a_{2,k}-a_{1,k})H}e^{it_kb_{2,k}\psi_0}e^{-ia_{2,k}H}.
\end{aligned}
\end{equation}
}
Here $X_0=[H,\psi_0]$ and $Y_0=i[H,X_0]$ are the fixed basis from Theorem~\ref{thm:inv-2d}. The real coordinates $(x_k,y_k)$ determine $(a_{1,k},a_{2,k},b_{1,k},b_{2,k})$ through \eqref{eq: 5 parameter}. If $G_k=0$, set $V_k=I$. The positive sign corresponds to maximizing $f$, equivalently minimizing $-f$.

Using a constant step size $t = 1/L_{\mathrm{Rie}}$, the Riemannian update achieves the targeted convergence rate. The final statement is given as follows.
\begin{theorem}[Baseline Complexity]\label{thm-bese-complexity}
Suppose we run the iteration \eqref{eq:r4-gradient-iteration} with the 5-factor retraction $\mathrm{R}_U$ defined in \eqref{eq: 5-factor}, and choose a fixed step size $t_k= 1 / L_{\mathrm{Rie}}$, where $L_{\mathrm{Rie}} = \mathcal{O}(\sqrt{N/M})$ is defined as above.
Then, for any $\varepsilon>0$,
\begin{equation*}
    T \geq \left\lceil\frac{2L_{\mathrm{Rie}}}{\varepsilon^2} \right\rceil \quad \Longrightarrow \quad \min _{0 \leq k \leq T-1}\left\|\operatorname{grad} f\left (U_k\right)\right\| \leq \varepsilon.
\end{equation*}
\end{theorem}
Thus the second stage produces a proof plan consisting of two main tasks. The first task is to establish a sharp estimate of $L_{\text{Rie}}$ on $\mathrm{U}(N)$ with the canonical metric and the 5-factor retraction. The second task is to verify that the assumptions of the Riemannian complexity theorem are satisfied by our algorithm and manifold. The remaining work, namely the refinement of inequalities, calibration of constants, and numerical stress tests for the bound on $L_{\text{Rie}}$, is carried out in the third stage.

\subsubsection{Parameter discovery: iterative estimates and numerical tests}\label{sec: constant}

Along the chosen route, some lemmas or inequalities involve parameters or coefficients that need sharp bounds. At this stage, human experts iterate with the AI on two parallel tracks.

(1) \textbf{Parameter proposals and goal checks.} Human experts state the explicit requirements that the final theorem imposes on the parameters. For example, that certain error terms must be bounded by a fraction of $\varepsilon$. The AI proposes candidate values or bounds.  Human experts test them against these requirements. Failed candidates are revised with short explanations (e.g. “this choice violates the inequality for large $N$”). Passing candidates are kept as the current best estimates. Over a few rounds, this yields a small table of admissible constants with clear justifications.

(2) \textbf{Numerical stress tests for inequalities.} For inequalities tied to the parameters, human experts ask the AI to propose simple low-dimensional checks and executable code. They execute the checks locally and verify whether each sample satisfies the inequality. When counterexamples appear, the model helps analyze the pattern, for instance whether violations occur near the boundary, and proposes tightened assumptions for human review.

By alternating these steps, human experts break the large problem into smaller, verifiable pieces that the AI can help address. The AI assists with variant generation, quick retrieval of known bounds, and parameter tuning, while human experts control the framework, proof logic, and acceptance criteria. At the end, the conjecture is refined into a stable theorem with a clear scope, a documented proof path, and constants checked by both analysis and targeted experiments.

\textbf{Example.}
Continuing the running example, to prove the Lipschitz estimate above we first isolate the contribution of the retraction. We focus on the following two standard inequalities for a retraction $\operatorname{Retr}(\cdot)$.
\[
    \left\|\operatorname{Retr}_x(\eta)-x\right\| \leq \alpha \|\eta\|,  \quad
    \left\|\operatorname{Retr}_x(\eta)-x-\eta\right\| \leq \beta \|\eta\|^2 .
\]
We ask the AI to propose candidate values for $(\alpha,\beta)$ that are valid on $\mathrm{U}(N)$ for our chosen five factor retraction $\mathrm{R}_U$ in Theorem \ref{thm: 5-factor}. The language model suggests
\(
    \alpha_{\text{tight}} = 1, \;\beta_{\text{tight}} = \frac{1}{4 c_0},
\)
where
\(
c_0 :=\left\|X_0\right\|_F=\left\|Y_0\right\|_F=\left\|\left[X_0, \psi_0\right]\right\|_F=\frac{\sqrt{2 M (N-M)}}{N},
\)
and flags these constants as potentially tight. Human experts adopt these values after symbolic verification. The processed theorem states as follows.
\begin{theorem}[Tight first and second order bounds]\label{thm-5factor-bounds}
Consider the 5-factor retraction $\mathrm{R}_U$ defined in \eqref{eq: 5-factor}. For all $U \in \mathrm{U} (N)$ and $\eta \in \mathcal{W} U$, it holds
\begin{equation*}
    \left\|\mathrm{R}_U (\eta)-U\right\|_F \leq\|\eta\|_F, \qquad\left\|\mathrm{R}_U (\eta)-U-\eta\right\|_F \leq \frac{1}{4 c_0}\|\eta\|_F^2.
\end{equation*}
Moreover, both estimates are globally tight: as $\eta \rightarrow 0$, the ratio in the first bound tends to $1$, and the ratio in the second bound tends to $1/(4c_0)$.
\end{theorem}
Guided by the prompt in Appendix~\ref{appendix: coeffi}, the language model is used within an explicit propose--test--prove loop. It proposes $(\alpha,\beta)$ and executable numerical stress tests for random unitary points and tangent vectors. Human authors run these tests locally and revise the candidate constants if violations or unstable behavior are detected. Numerical evidence identifies promising constants. The final theorem relies on the subsequent symbolic proof and tightness argument.

\subsubsection{Refining the theorem through structural insights and improved results}\label{sec: refining the theorem}
After the first successful proof, the collaboration enters a refinement stage. In the final stage, human experts review the baseline result to identify structural properties of the problem that may have been underutilized in the initial rough analysis. The goal here is to strengthen the theorem, including improving convergence rates, weakening assumptions, or simplifying the final constants.

In this phase, the interaction loop shifts from verification to discovery of structure. Human experts hypothesize stronger conditions (such as strong convexity, or the Polyak-Łojasiewicz (PL) inequality) and ask the AI to attempt to derive these conditions for the scoped object. If such a bound relating the gradient norm to the function value gap can be verified, the complexity analysis can be upgraded without changing the underlying algorithm. This stage transforms a generic ``worst-case'' theorem into a ``problem-specific'' sharp result.

\begin{table}[!t]
\centering
\footnotesize
\manualtablecaption{tab:rate-comparison}{Convergence guarantees before and after the PL refinement.}
\begin{tabular}{>{\raggedright\arraybackslash}p{0.24\linewidth}>{\raggedright\arraybackslash}p{0.32\linewidth}>{\raggedright\arraybackslash}p{0.32\linewidth}}
\hline
Analysis stage & Assumption used & Iteration guarantee \\
\hline
Smoothness baseline & Riemannian pullback-gradient smoothness with $L_{\mathrm{Rie}}$ & $O(L_{\mathrm{Rie}}\varepsilon^{-2})$ iterations for $\min_k\|\operatorname{grad} f(U_k)\|\leq\varepsilon$ \\
PL refinement & Iterates already in and remaining in the local PL and smoothness region, with $L_{\mathrm{Rie}}=O(\sqrt N)$ & $O(L_{\mathrm{Rie}}\log(1/\varepsilon))=O(\sqrt N\log(1/\varepsilon))$ iterations for $1-q_T\leq\varepsilon$ \\
\hline
\end{tabular}
\end{table}

\textbf{Example.} Returning to our running example, Theorem \ref{thm-bese-complexity} establishes a convergence rate of $O(\varepsilon^{-2})$. This is the standard guarantee for Riemannian gradient descent on general nonconvex smooth functions. Human experts then examine whether the specific objective function $f(U) = \operatorname{Tr}(H U \psi_0 U^\dagger)$ has better local geometry near the optimum, specifically a relationship between the magnitude of the gradient and the suboptimality gap. We ask the language model to search for structural mechanisms that could sharpen the convergence result. The concrete prompt is given in Appendix \ref{appendix: refining the theorem}.

The model identifies the PL property as a plausible route and lists the verification steps required to test this approach. Human experts then check these steps and formulate the refined result below.

\begin{theorem}[Best Complexity via PL Inequality]\label{thm-best-complexity}
Consider the iteration \eqref{eq:r4-gradient-iteration} under the same assumptions as in Theorem~\ref{thm-bese-complexity}. Let $q_k:=f\left(U_k\right) \in[0,1]$ be the objective value at step $k$, and let $q_{\max} = 1$ be the global maximum.

Assume that the initial iterate lies in a region where the pullback-gradient smoothness bound with constant $L_{\mathrm{Rie}}$ and the following PL inequality hold. Assume further that all subsequent iterates $U_k$, for $1\leq k<T$, remain in this region. The theorem does not establish entry into the region from an arbitrary initialization.
\begin{equation*}
    \left\|\operatorname{grad} f(U_k)\right\|^2 \geq \frac{1}{3}\left(1-q_k\right), \qquad 0\leq k<T.
\end{equation*}
Then, for any target accuracy $0<\varepsilon \leq 1-q_0$, the iterates satisfy the optimality gap $1- q_T \leq \varepsilon$ in at most
$T=\left\lceil 6L_{\mathrm{Rie}} \log \left(\frac{1}{\varepsilon}\right) \right\rceil$
iterations.
\end{theorem}

\begin{wrapfigure}{r}{0.46\textwidth}
    \centering
    \setlength{\abovecaptionskip}{4pt}
    \includegraphics[width=\linewidth]{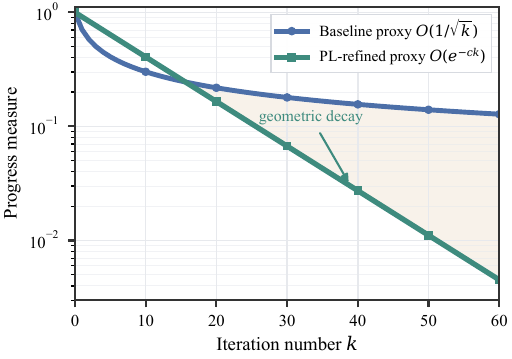}
    \caption{PL-refinement summary.}
    \label{fig:rate-comparison}
\end{wrapfigure}

This final improvement illustrates the role of the refinement stage. The baseline proof provided the smoothness-based guarantee, while the later structural analysis used the specific geometry of the objective to prove a local PL inequality and combine it with \(L_{\mathrm{Rie}}=O(\sqrt N)\). Table~\ref{tab:rate-comparison} summarizes the refinement. The proof is checked by human experts within the workflow described here.

\section{Roles of human and AI} \label{sec:roles}

In the workflow studied here, human authors control the mathematical specification. They fix the objects, assumptions, admissible operations, and criteria for accepting a statement. The LLM works within this specification by proposing candidate proof routes, lemmas, counterexamples, constructions, and computational checks. In the Grover case study, human authors prune these proposals using the Grover plane, projector identities, scalar recursions, endpoint regimes, and small matrix tests. Proposals that conflict with these checks are repaired or rejected.

Selected prompt records and failure records make this division of labor observable. Together, these materials show the tasks posed to the model, representative model suggestions, the checks applied by the human authors, and the resulting acceptance, repair, or rejection decisions. In the second, related multi-oracle case study, rejected routes included an intersection oracle outside the stated gate model, updates that no longer represented the average Riemannian gradient, and convention errors in matrix products. The prompt structure supports the checking process by asking the model to restate assumptions, separate proof attempts from counterexample searches, and expose dependencies.

Human authors retain full ethical and academic responsibility for the results and for the scholarly record. The workflow is primarily a natural-language proof workflow, supported by prompt records, the Python script for checks, and partial auxiliary formalization in the public repository. Proof assistants can provide stronger certification after definitions and theorem statements stabilize. They are complementary to the workflow studied here. An AI-generated suggestion enters the manuscript only after human authors verify the assumptions, repair gaps, rewrite the argument, and decide that the claim belongs in the final record. The workflow presupposes domain experts who can identify shifted assumptions, notation drift, missing arguments, and misuse of standard lemmas.

\section{Conclusion} \label{sec: conclusion}
We presented a human-in-the-loop workflow for mathematical research. In this workflow, human experts set the mathematical scope, check the model's suggestions, and decide which statements enter the paper. The LLM is used to explore possible proofs, counterexamples, and auxiliary claims inside this human-defined frame. In the case study, the interaction produces checkable records and helps organize the refinement from a generic smooth \(O(L_{\mathrm{Rie}}\varepsilon^{-2})\)-type guarantee to a PL-based \(O(\sqrt N\log(1/\varepsilon))\) complexity bound for the Grover-compatible Riemannian gradient method under the assumption that the iterates already lie in and remain in the region where the PL inequality and pullback-gradient smoothness bound hold.  These materials provide a concrete reporting format for AI-assisted mathematical work. They document what the model proposed, how each proposal was verified, which routes were rejected or corrected, and which validated statements were ultimately incorporated into the final paper.

\section*{Acknowledgments}

The authors used AI-based tools for English-language polishing. The authors are grateful to the (associate) editor and the anonymous referees for their detailed and valuable comments and suggestions.

\medskip

\textbf{Funding.}
This work was supported in part by the National Natural Science Foundation of China under the grant numbers 12501419, 12331010 and 12288101, and National Key Research and Development Program of China under the grant number 2024YFA1012903.

\smallskip

\textbf{Availability of materials.}
The materials supporting this study are available in the public GitHub repository at
\url{https://github.com/optsuite/MathResearchPrompts}.

\medskip

\textbf{Declarations}

\smallskip

\textbf{Conflict of interest.}
The authors have no relevant financial interests to disclose.

\appendix
\section{Grover-related supplementary case study and prompt records}\label{appendix}
This appendix presents the related multi-oracle exploration, representative rejected routes, the corresponding qualitative failure records, and selected prompts for the main Grover-compatible case study. The versions printed here are shortened for readability. The repository at \url{https://github.com/optsuite/MathResearchPrompts} provides the prompt--response materials, reusable templates, finite-difference retraction checks, and auxiliary formalization code. The reusable templates include slots for domain-specific assumptions, admissible operations, boundary cases, and verification checks.

\paragraph{Computational environment}
The interactions used in the main case study were conducted with GPT-5.1 through the web-based chat interface, accessed between September 2025 and December 2025. No local LLM deployment and no local GPU inference were used. The numerical checks reported in the main case study were small CPU-based matrix tests which were run on a MacBook Pro with an Apple M4 chip, a 10-core CPU, 24 GB memory, and Python 3.10.

\subsection{Multi-oracle case study}\label{appendix:multi-oracle-exploration}

This second, related multi-oracle case study keeps the Grover-compatible viewpoint of the main case study but replaces one oracle projector by several projectors. It lies outside the main theorem sequence and examines which parts of the workflow transfer and which parts require new checks. The starting point is the averaged objective
\[
    f_{\mathrm{avg}}(U)=\frac{1}{r}\sum_{j=1}^{r}\operatorname{Tr}(H_jU\psi_0U^\dagger),
\]
where \(H_1,\ldots,H_r\) are oracle projectors. If a state satisfies all oracle conditions, then each score \(\operatorname{Tr}(H_jU\psi_0U^\dagger)\) should be close to one. Maximizing \(f_{\mathrm{avg}}\) is therefore a natural way to express simultaneous progress toward all oracle conditions.

In the commuting-projector setting, the model suggested using joint membership patterns. A pattern \(b\in\{0,1\}^r\) records, for each \(j\), whether a component lies in the range of \(H_j\). It defines a corresponding pattern projector \(E_b\). This gives a concrete reduction. The relevant state components lie in the span of the nonzero vectors \(E_b|\psi_0\rangle\), whose dimension is at most \(2^r\). In this pattern basis, both \(f_{\mathrm{avg}}\) and the corresponding average-gradient direction become finite-dimensional objects that can be compared with full matrix simulations. This reduction serves as checked exploratory material supplementing the theorem sequence in the main case study.

For $r=2$, we additionally record a frame check in the fully active case. Let
{
\[
\mathcal G
=
\{(0,0),(0,1),(0,2),(1,0),(1,1),(1,2),(2,0)\},
\]}
and, for $g=(g_1,g_2)\in\mathcal G$ and $\epsilon\in\{0,1\}$, define
{
\[
W_{g,\epsilon}
=
e^{\frac{\pi i}{2}g_1H_1}
e^{\frac{\pi i}{2}g_2H_2}
e^{\frac{\pi i}{2}\psi_0}
e^{\frac{\pi i}{4}\epsilon H_1}
e^{\frac{\pi i}{4}\epsilon H_2},
\qquad
P_{g,\epsilon}=W_{g,\epsilon}\psi_0W_{g,\epsilon}^{\dagger}.
\]}
Throughout this construction, matrix products follow the standard convention that the rightmost factor acts first. Thus the construction contains $7\times2=14$ projectors generated only from the admissible phase gates associated with $H_1,H_2$, and $\psi_0$. For a complex matrix $A$, define its real vectorization by
{
\[
\operatorname{vec}_{\mathbb R}(A)
:=
\begin{bmatrix}
\operatorname{Re}\operatorname{vec}(A)\\
\operatorname{Im}\operatorname{vec}(A)
\end{bmatrix}.
\]}
We define $B_{\mathrm{det}}$ as the real matrix whose columns are the vectorized deterministic frame directions,
{
\[
B_{\mathrm{det}}
:=
\bigl[
\operatorname{vec}_{\mathbb R}(iP_{g,\epsilon})
\bigr]_{(g,\epsilon)\in\mathcal G\times\{0,1\}}.
\]}
Let $e_b$, $b\in\{00,10,01,11\}$, denote the canonical basis vectors of the fully active four-pattern representation, and write $|b|:=b_1+b_2$. Order the patterns as $00<10<01<11$. For every pair $b<c$ with $|b|\neq|c|$, define
{
\[
D^{(1)}_{bc}:=e_be_c^\ast-e_ce_b^\ast,
\qquad
D^{(2)}_{bc}:=i(e_be_c^\ast+e_ce_b^\ast).
\]}
These matrices form a real basis for the average-gradient subspace in the fully active two-oracle pattern representation. We define $B_{\mathrm{avg}}$ by taking their real vectorizations as columns,
{
\[
B_{\mathrm{avg}}
:=
\bigl[
\operatorname{vec}_{\mathbb R}(D^{(1)}_{bc}),
\operatorname{vec}_{\mathbb R}(D^{(2)}_{bc})
\bigr]_{b<c,\,|b|\neq|c|}.
\]}
The frame condition is tested by asking whether the columns of $B_{\mathrm{avg}}$ lie in the real column span of $B_{\mathrm{det}}$. In the five deterministic test configurations and 200 additional random positive pattern-weight configurations, we obtained
{
\[
\operatorname{rank}(B_{\mathrm{det}})
=
\operatorname{rank}[B_{\mathrm{det}}\ B_{\mathrm{avg}}]
=14,
\]}
with the largest least-squares projection residual equal to $r_{\max}=2.029\times10^{-13}$. The reduced pattern-basis and corresponding full simulations agreed with a maximum state-vector discrepancy of $2.091\times10^{-15}$. For $N\in\{64,128,\ldots,16384\}$, fitting the observed iteration counts to $T_N=a\sqrt N+b$ gave $T_N=3.245224\sqrt N-21.727607$ with $R^2=0.983305$. Least-squares linear regression of $\log T_N$ on $\log N$, using the observed iteration counts at the same values of $N$, gave a log--log slope of $0.542181$. These computations provide numerical diagnostics for the two-oracle construction within the tested configurations and problem sizes.

Several components of the workflow transfer directly: specification of admissible operations, generation of candidate structural reductions, and numerical comparison with full matrix dynamics. The multi-oracle setting additionally requires checks that are absent or less prominent in the single-oracle case, including gate-model admissibility, consistency with the average Riemannian gradient, and product-order conventions.

The second case study also supplied negative evidence. Several natural-looking routes solved a different mathematical problem, lost the required gradient structure, or depended on implementation conventions that the model left implicit. The next subsection records a representative chain of such failures.

\subsection{Representative failed routes from the second multi-oracle case study}\label{appendix:negative-example}

We retain four representative failures from this case study to make the failure record concrete: an inadmissible intersection oracle, a direction incompatible with the average gradient, a matrix-product convention error, and overstrong numerical convergence language. Consider again the multi-oracle objective
\[
    f_{\mathrm{avg}}(U)=\frac{1}{r}\sum_{j=1}^{r}\operatorname{Tr}(H_jU\psi_0U^\dagger),
\]
where \(H_1,\ldots,H_r\) are oracle projectors. The intended gate model permits phase gates generated by \(\psi_0\) and by the individual \(H_j\)'s, so admissible updates are built from these original gates. A phase gate encoding the intersection of all marked subspaces lies outside this model.

A first model-generated route violated this rule. In the commuting case, the model proposed replacing the average objective by an intersection projector such as
\[
    H_{\cap}=H_1H_2\cdots H_r
\]
and then optimizing
\[
    \operatorname{Tr}(H_{\cap}U\psi_0U^\dagger).
\]
This is mathematically natural, but it assumes access to the phase gate \(e^{-itH_{\cap}}\). That gate already encodes the intersection and therefore changes the gate model, bypassing the intended construction from admissible gates. Human experts rejected this route and tightened the prompt to require that every proposed update be generated from \(\psi_0,H_1,\ldots,H_r\) themselves.

A second failure was more subtle. Writing \(\rho_U=U\psi_0U^\dagger\), and using the Hermitian-generator convention in which the infinitesimal update is written with \(e^{-itG}\), the average objective has generator
\[
    G_{\mathrm{avg}}(U)=\frac{i}{r}\sum_{j=1}^{r}[\rho_U,H_j].
\]
Some proposed updates used one component direction \(i[\rho_U,H_j]\) at a time. Such an update may be generated from admissible gates and may increase the single score \(\operatorname{Tr}(H_j\rho_U)\). It can still fail as an update for \(f_{\mathrm{avg}}\), because the average-gradient direction depends on all commutators in the sum above. In numerical tests, related projected-gradient variants stalled away from the intersection after the projection removed components of \(G_{\mathrm{avg}}(U)\) needed for progress. A valid proposal therefore needs both admissible gate generation and compatibility with the Riemannian gradient of the stated objective.

The later frame-compatible route arose from these failures. It represents the average-gradient direction by projectors generated from admissible gates and preserves the original gate model. The implementation still required human correction. In one tested frame construction, the frame condition failed when the product word was interpreted as a left-to-right action on states. It held under the standard matrix-product convention that the rightmost factor acts first. The same case study also led us to weaken numerical convergence language. Experiments suggested \(f_{\mathrm{avg}}(U_t)\to1\) in tested cases, while the analysis justified stationarity, for example \(\|G_{\mathrm{avg}}(U_t)\|\to0\), under the stated assumptions. This example shows why the workflow records rejected routes, implementation conventions, and overstrong numerical interpretations alongside the retained routes in this case study.

The recovered record also shows that substantial human correction was not limited to rejecting isolated false claims. An early formulation relied on dynamical-Lie-algebra language that was mathematically relevant but poorly aligned with the intended optimization presentation, and it was later removed from the algorithmic main line. Likewise, a frame condition that was initially supported by random constructions was replaced by an explicit deterministic frame only after the authors required a fixed construction and a direct span check. These episodes distinguish a discarded route from a salvageable response whose central idea survives after a change of language, assumptions, or verification standard.

\begin{table}[htbp]
\centering
\scriptsize
\renewcommand{\arraystretch}{1.16}
\manualtablecaption{tab:failure-modes}{Failure modes documented in the second multi-oracle case study.}
\begin{tabularx}{0.98\linewidth}{>{\raggedright\arraybackslash}p{0.16\linewidth}>{\raggedright\arraybackslash}p{0.21\linewidth}>{\raggedright\arraybackslash}p{0.25\linewidth}X}
\hline
Failure mode & Observable symptom & Required check & Disposition \\
\hline
Gate-model drift & The intersection projector requires a phase gate absent from the stated gate model. & Trace every proposed update to gates generated by \(\psi_0,H_1,\ldots,H_r\). & Reject the intersection-oracle route and enforce the admissible gate list. \\
Average-gradient mismatch & An update follows one oracle direction or loses components required for average progress. & Compare the proposed generator with \(G_{\mathrm{avg}}(U)\). & Reject or revise the update until it matches the stated objective's Riemannian gradient. \\
Convention ambiguity & The frame condition changes with the interpretation of product order. & Declare that the rightmost factor acts first and rerun the check. & Correct the implementation and record the failed convention. \\
Numerical overinterpretation & Tests suggest \(f_{\mathrm{avg}}(U_t)\to1\), while the analysis establishes stationarity. & Separate observed convergence from the proved statement. & State stationarity and label convergence as a tested observation. \\
\hline
\end{tabularx}
\end{table}

The table summarizes the four failure modes documented above and records the corresponding symptom, check, and disposition. These cases were deliberately selected to illustrate distinct failure mechanisms. They are not an exhaustive or randomly sampled set of model outputs and therefore do not provide a base rate for correct, salvageable, or unusable responses. We make no frequency or general-performance inference from these four examples.

\subsection{Prompt records for the main Grover-compatible case study}\label{appendix:prompt-records}

The remaining appendix entries index the prompt families used in the main Grover-compatible case study.

\begin{table}[htbp]
\centering
\manualtablecaption{tab:prompt-index}{Prompt index by intent.}
\begin{tabular}{p{0.12\linewidth}p{0.28\linewidth}p{0.45\linewidth}}
\hline
Appendix label & Intent & Main verification check \\
\hline
\ref{appendix: early} & Scoping and topic conceptualization & Human pruning of overbroad or ill-posed directions \\
\ref{appendix: sharpened claims} & Grover route framing & Alignment checks against known objects and symmetries \\
\ref{appendix:direct proofs} & Direct proof with fixed target & Line-by-line proof verification and no fabricated references \\
\ref{appendix: uncertain} & Prove-or-disprove analysis & Parallel proof and counterexample branches \\
\ref{appendix: proposal} & Candidate theorem discovery & Edge cases, numerical probes, and selection of surviving claims \\
\ref{appendix: human abstraction} & Cross-setting transfer & Reuse only invariant structures that survive checks in the new setting \\
\ref{appendix: property constrained} & Product retraction synthesis & Finite-difference velocity tests and symbolic proof of (P1)--(P3) \\
\ref{appendix: scoping} & Statement specification & Explicit objects, ambient space, metric, and assumptions \\
\ref{appendix: route} & Route confirmation & Literature check and replacement of ambient Euclidean reasoning by Riemannian reasoning \\
\ref{appendix: coeffi} & Constant discovery & Numerical stress tests followed by symbolic bounds \\
\ref{appendix: refining the theorem} & Rate refinement & Verification of the PL inequality and resulting iteration count \\
\hline
\end{tabular}
\end{table}

\EnableCompactMode

\subsubsection{Prompts for Section \ref{sec: early-stage}}\label{appendix: early}

You are an expert in both quantum algorithms and mathematical optimization.

I want to identify promising points of contact between quantum algorithms and optimization that could lead to clean, analyzable research problems. You can refer to the paper I give you. Please do the following:

\begin{enumerate}
    \item Give a concise briefing that figures out:
    \begin{itemize}
        \item The main quantum algorithmic primitives that are relevant here.
        \item The main optimization paradigms that might interact with them.
        \item Typical objectives, loss functions, and structural assumptions on each side.
    \end{itemize}
    \item Propose 5-8 concrete candidate research directions at this interface. For each direction, provide:
    \begin{itemize}
        \item A short description of the core idea.
        \item The main mathematical objects and assumptions.
        \item The nearest classical or existing results and what gap or limitation they leave.
        \item A plausible class of analysis techniques.
    \end{itemize}
    \item For each direction, add 2-3 sentences of feasibility notes:
    \begin{itemize}
        \item Why would this direction admit a clean theorem (or why it might fail).
        \item Any obvious degenerate or trivial regimes to watch out for.
        \item Whether you expect complexity or convergence guarantees to be within reach.
    \end{itemize}
\end{enumerate}

Format your answer as a numbered list of candidate directions, each with compact bullet points and feasibility notes. Avoid overly speculative ideas that are disconnected from existing theory.

\subsubsection{Prompts for Section \ref{sec: concrete-idea}}\label{appendix: sharpened claims}

You are a theoretical mathematician specializing in the intersection of quantum algorithms and Riemannian optimization. Your goal is to build connection between quantum algorithmic structures and Riemannian optimization.
We are investigating a Riemannian optimization explanation for quantum search algorithm. Specifically, we hypothesize that Grover's algorithm can be rigorously interpreted as a form of Riemannian gradient descent. We need to identify the underlying connection before attempting a proof.

\textit{Task:} Please provide a detailed theoretical framing for this intersection:

\begin{enumerate}
    \item {Object Identification:} Define the precise mathematical objects on both sides. This must include the manifold of quantum states and the specific Riemannian gradient update rules that serve as the continuous analog.
   \item {Properties and Symmetries:} Analyze the essential properties of Grover iterations, such as unitarity and query complexity. How do these map to the oracle complexity in Riemannian optimization?
   \item {Alignment Check:} Identify specific regimes or limiting cases where the standard gradient like updates and the discrete Grover iterations coincide structurally.

\end{enumerate}

\subsubsection{Prompts for Section \ref{sec:direct proofs}}\label{appendix:direct proofs}
The following direct-proof prompt already contains the no-fabricated-references rule. This rule is highlighted here because it functions as a methodological safeguard. The model may invoke standard named results, but citation status and bibliographic support remain subject to human verification before any claim is incorporated into the manuscript.

You are a careful mathematical assistant collaborating with human experts.

When you are asked to prove a statement, you must:

\begin{enumerate}
\item restate all assumptions, notation, and domains of all variables.

\item give a numbered, line-by-line argument in which every nontrivial inference is explicitly justified.

\item cite standard results by their usual names (e.g.\ spectral theorem, Cauchy--Schwarz inequality) and never fabricate new references.
\end{enumerate}

Prove the following statement in a concise but fully rigorous way. In your proof, you should in particular:

\begin{enumerate}
\item explain carefully why $\operatorname{grad} f(U)=0$ is equivalent to $[H,\psi_U]=0$.

\item prove both directions of the optimality claim: that $\operatorname{grad} f(U)=0$ implies that $U$ is a global optimizer with $f(U)\in\{0,1\}$, and that any global optimizer must satisfy $\operatorname{grad} f(U)=0$.
\end{enumerate}
The statement gives as [Statement of Theorem \ref{thm: optimality}].

\textit{Output format.}
Return only:

\begin{enumerate}
  \item a LaTeX \texttt{proof} environment with a numbered, step-by-step derivation.
  \item a short LaTeX \texttt{itemize} list of adversarial checks a reader could perform.
\end{enumerate}
Do not output any text outside these two parts.

\subsubsection{Prompts for Section \ref{sec: clear target}}\label{appendix: uncertain}

\textit{Content.}
You are a careful mathematical assistant participating in a \emph{prove-or-disprove}
workflow. Your task is to analyze the structural statement below about commutators
of Hermitian matrices and the Lie algebra $\mathfrak{su}(n)$.
When reasoning about this statement, you must:

\begin{enumerate}
\item Restate all assumptions, notation, and domains of the variables
        (including the ambient space $M_n(\mathbb{C})$, (skew-)Hermitian conditions, and the trace operator).

\item Run \emph{two} explicit branches:

\begin{itemize}
    \item {Proof branch:} Try to show that every traceless
          skew-Hermitian matrix can be written as a commutator $[A,B]$
          with $A^\dagger = A$ and $B^\dagger = B$. Use concrete
          constructions, basis expansions, or dimension arguments.
    \item {Disproof branch:} In parallel, look for structural
          obstructions or counterexamples. Analyze the subspace
          spanned by such commutators, check invariants, and probe whether some elements of $\mathfrak{su}(n)$ might fail to be commutators.
\end{itemize}

\item Within each branch, give a line-by-line argument in which every non-trivial inference is explicitly justified.
\item Cite standard facts (e.g.\ basic properties of commutators,
        spectral theorem, Lie algebra identities) by their usual names
        and never fabricate new references.
\item  After exploring both branches, provide a \emph{final verdict}:
        decide whether the proposition is true or false. If it is true,
        present a concise but fully rigorous proof. If it is false,
        present an explicit counterexample and explain why it fails.
\end{enumerate}

\textit{Target statement (to prove or disprove).}
[Statement of Proposition \ref{prop: uncertain}].

\textit{Required output format.}
Return only the following three parts in LaTeX:

\begin{enumerate}
    \item A {Proof branch} subsection with a numbered, step-by-step
        derivation attempting to prove the proposition.
    \item  A {Disproof branch} subsection with a numbered, step-by-step analysis searching for obstructions or counterexamples.
    \item A {Final verdict} paragraph stating clearly whether the
        proposition is true or false, together with either a complete proof
        or a concrete counterexample, followed by a short {itemize}
        list of adversarial checks a human reader could perform.
\end{enumerate}
Do not output any commentary outside these three parts.

\subsubsection{Model reply for Section \ref{sec: rich proofs}}\label{appendix: rich}

The part of the model output that inspired us to obtain Theorem \ref{thm: aha} is as follows.

\textit{The Invariant ``Jordan'' Plane (Working Subspace).}
Because $H$ is a projector and $\psi_0$ is rank-one, the evolution is strictly confined to the Krylov subspace $\mathcal{K} = \text{span}\{ |\psi_0\rangle, H|\psi_0\rangle \}$.

\begin{enumerate}
    \item \textit{Algebraic reason:} The operators $H$ and $\psi_0$, when restricted to the subspace containing $|\psi_0\rangle$, form a closed $2\times 2$ algebra (a single Jordan block of the pair $\{H, \psi_0\}$).
    \item \textit{Geometric intuition:} This is a ``Grover plane'' on the projective Hilbert space connecting the start state $\psi_0$ and the target state (the projection of $H$).
    \item \textit{Subspace:} The state $\psi_k$ remains in the same 2D ``Jordan plane'' $\mathcal{K}$ as the baseline. The operators $H$ and $\psi_0$ act as $2\times 2$ matrices on this subspace (specifically, rank-1 projectors with overlap angle $\theta_0 = \arccos \sqrt{q_0}$).
    \item \textit{Dynamics:} While the baseline flows continuously along the gradient $[H, \psi_k]$, the two-factor product ansatz performs discrete rotations within the same 2D plane. It acts as a ``bang-bang'' control to approximate the geodesic flow.
\end{enumerate}

\textit{Our observation} The model identified that the system's evolution is invariant under a specific 2D subspace, effectively reducing the dynamics to SU(2) rotations. We realized that this geometric confinement implies the existence of a low-dimensional algebraic structure. Consequently, we were able to derive Theorem \ref{thm: aha}, which replaces high-dimensional matrix operations with efficient, closed-form scalar recurrences in $(x_k,y_k,q_k)$ suitable for classical tracking.

\subsubsection{Prompts for Section \ref{sec: AI-guided proposal}}\label{appendix: proposal}

You are a careful mathematical assistant helping with theorem discovery for a fixed dynamical setting. The assumptions and notation are fixed. The conclusions are not.

\textit{Setting.} [Theorem setting for Proposition \ref{thm:exp-update}].
We study the following update rules for $U_{k+1}$ and $\psi_{k+1}$.

\textit{Exact exponential step (``Exp update''):}  [exact update scheme in Proposition \ref{thm:exp-update}].

Our broader goal is to surface structurally meaningful consequences of this dynamics. These may include (but are not limited to) invariants, norm and energy identities, dimensional reductions, stability or monotonicity statements, convergence and one shot optimality conditions, and useful lemmas for later analysis.

\textit{Task.}
Starting from the setting above, use assumption-driven discovery to propose
candidate conclusions. Proceed as follows.

\begin{enumerate}
    \item {Restate} all assumptions and domains (projector properties of $H$, rank-one nature of $\psi_k$, ranges of $q_k$ and $\gamma_k$, and the unitary dynamics for $U_k$ and $\psi_k$).
    \item {Generate candidate statements.} Propose a list of concise,
        LaTeX-ready candidate claims. For each candidate:

\begin{itemize}
    \item assign a \emph{type label} that you choose (for example,
\emph{Invariant}, \emph{Norm identity}, \emph{Scalar recursion},
\emph{Convergence guarantee}, etc.).
    \item provide a one or two sentence informal explanation of what the
claim says and why it might be useful.
\end{itemize}
At least some candidates should reflect both \emph{local} information (e.g.\ behavior in the Grover plane,) and \emph{global} information (e.g.\ long-time dynamics), but you are free to propose additional categories of structural statements beyond these.
\item {Selection.} From the candidates that pass the tests, select a coherent subset (typically $3$-$5$ items) that best captures the core structure of the Exp update. Combine them into a single proposition with numbered parts (i), (ii), (iii).
\item {Clarity and rigor.} When presenting the final proposition, give short, line-by-line derivations for each numbered item, with explicit justifications and citations of any standard facts used (e.g.\ properties of projectors, commutators, unitary rotations in a two-dimensional subspace). Avoid hand-waving and do not fabricate references.
\end{enumerate}

\textit{Required output format.}
Return only:
\begin{itemize}
    \item a numbered list of candidate statements with their type labels, informal explanations, and pass/fail/inconclusive labels from the light-weight tests.
    \item a final LaTeX proposition environment collecting the selected statements into a single theorem-style statement, together with a concise, step-by-step proof.
    \item a short {itemize} list of adversarial checks that a human reader could apply to stress-test the final proposition and its assumptions.
\end{itemize}

\subsubsection{Prompts for Section \ref{sec: human abstraction}}\label{appendix: human abstraction}

You are a careful mathematical assistant collaborating with human experts on
theorem discovery and \textit{cross-setting transfer}.

Treat the setting below as one instance of a more general pattern rather than as an isolated calculation. Your goals are extract a reusable schema that can be instantiated in nearby settings and apply it on a similar setting.

\textit{Baseline target.}
[Statement of Proposition \ref{thm:exp-update}].

\textit{Instructions.}
Organize your answer into three sections.

\begin{enumerate}
    \item {Abstract schema}. Extract the structural ingredients behind the baseline target including but not limited to.
\begin{itemize}
    \item the working invariant subspace.
    \item the associated “gradient plane” or tangent plane.
    \item the scalar progress coordinate.
    \item the structured local update
\end{itemize}
\item  {Transfer notes} (invariant-product setting).

Transfer the above conclusion in the following setting. [Basic setting in Theorem \ref{thm:inv-2d}]. In this product setting:
\begin{itemize}
    \item Propose candidate analogues of the baseline schema.
    \item Formulate concrete transferred statements (propositions or conjectures).
    \item Give me the proof of the new results you propose.
\end{itemize}
\end{enumerate}

\textit{Output format.}
a final LaTeX proposition environment collecting the selected surviving statements into a single theorem-style statement, together with a concise, step-by-step proof.

\subsubsection{Prompts for Section \ref{sec: synthesis}}\label{appendix: property constrained}

\textit{Notation note.} The curve notation in this template is harmonized with Theorem~\ref{thm: 5-factor}: $V(t;x,y)$ denotes the gate product, and $\gamma_U(t;x,y)=V(t;x,y)U$ denotes the retraction curve through $U$.

\textit{Content.}
You are a careful mathematical assistant for mathematical discovery.
The goal is to construct a short gate product $V(t;x,y)$ that satisfies
a fixed list of properties. The admissible building blocks and the target
properties are given, but the concrete product form is not.

\textit{Setting.}
Let $H=H^\dagger=H^2$ be a projector and let $\psi_0$ be a rank-one projector.
Define
\(
  X_0 := [H,\psi_0],\; Y_0 := i[H,X_0],\;
  \mathcal W := \mathrm{span}_\mathbb{R}\{X_0,Y_0\}.
\)
For each $(x,y)\in\mathbb R^2$, we consider tangent targets
\(
  Z = x X_0 + y Y_0 \in \mathcal W,
\)
and search for gate-product curves $V(t;x,y)$ on the unitary group with the
following properties: [Constraints (P1) - (P3)].

\textit{Task.}
Design $V(t;x,y)$ by property-constrained synthesis. Proceed as follows.
\begin{enumerate}
    \item {Restate the constraints.}
        Summarize the admissible factor types, the target space $\mathcal W$,
        and properties \textup{(P1)}–\textup{(P3)} in your own words, making
        clear which quantities are fixed data ($H,\psi_0,X_0,Y_0$) and which are design choices (product length, coefficient
        couplings, dependence on $t,x,y$).
    \item {Propose candidate product forms.}
        Generate a small, ordered list of candidate ansätze for $V(t;x,y)$:
    \begin{itemize}
        \item  Each candidate must be an explicit product of exponentials
                $e^{i\theta_j H}$ or $e^{i\theta_j\psi_0}$ with scalar
                parameters $(a_j,b_j,\dots)$ that may depend on $t,x,y$.
        \item  Prefer shorter products. Include candidates of different lengths,
                including length $3$, length $4$, length $5$, and longer products,
                so that there is a chance to find the shortest viable form.
        \item For each candidate, give a one- or two-sentence explanation of
                how the chosen ordering and parameter couplings might help
                satisfy \textup{(P2)}-\textup{(P3)}.
    \end{itemize}
    \item {Numerical finite-difference tests.}
        For each candidate, use a seed set $\mathcal S\subset\mathbb R^2$, a small step size $h>0$, and a tolerance $\varepsilon>0$ to write concrete code (e.g.\ in Python) for the following checks.
    \begin{itemize}
        \item implements $V(t;x,y)$ for given $(x,y)$ and parameter
                values.
        \item checks \textup{(P1)} syntactically.
        \item checks \textup{(P2)} exactly at $t=0$.
        \item checks \textup{(P3)} numerically on the seed set $\mathcal S$
                via the finite-difference test
                [Equation \eqref{eq: test}]
                and reports pass/fail outcomes for the test,
                including edge cases such as $(x,0)$ and $(0,y)$.
    \end{itemize}
    The code must be executable, and you should run it to verify your construction.
    \item {Selection and refinement.}
        Among all candidates that pass the numerical tests on $\mathcal S$,
        identify the shortest product that appears robust.
        For this selected candidate:
    \begin{itemize}
        \item summarize the final parameter constraints.
        \item restate the product in a clean, theorem-ready form.
        \item  sketch a proof plan for why \textup{(P1)}-\textup{(P3)} hold
                exactly, indicating which algebraic identities and expansions
                need to be verified.
    \end{itemize}
\end{enumerate}

\textit{Required output format.}
Return only:
\begin{enumerate}
    \item a numbered list of candidate product forms with their informal
             explanations and symbolic viability labels.
    \item the fully specified, shortest selected candidate in LaTeX, together with a brief, step-by-step derivation of the parameter constraints ensuring \textup{(P1)}-\textup{(P3)}.
    \item executable Python code for the numerical tests on $\mathcal S$,
             including the finite-difference check above, and a short
             {itemize} list of adversarial checks a human reader could
             apply to stress-test the construction.
\end{enumerate}

\subsubsection{Prompts for Section \ref{sec: scoping}}\label{appendix: scoping}

You are an expert mathematical research collaborator specializing in geometric analysis, numerical optimization, and precise theorem formulation. I will present a rough, informal target theorem or conjecture. Your goal is to help me make this statement precise and analyze its provability.

Core Responsibilities:

\begin{enumerate}
    \item {Clarification:} take my vague concepts and propose precise mathematical definitions. Explicitly define the ambient space, the metric, and the notion of gradient. List necessary assumptions required to make the statement well-posed.
    \item {Feasibility and proof strategy:} analyze the logical structure of the conjectured bound or equality. Propose feasible proof paths: Suggest specific techniques. Identify potential bottlenecks or technical difficulties in the proof.
    \item {Literature and context:}  connect the conjecture to standard results or known theorems in the field. If the conjecture seems standard or trivial, point it out. If it contradicts known counterexamples, warn me immediately.
\end{enumerate}

\textit{Format:} Use LaTeX for all mathematical expressions. Your response must be logically organized, prioritizing mathematical clarity and insight over formatting.

\subsubsection{Prompts for Section \ref{sec: route}}\label{appendix: route}
You are familiar with Riemannian optimization and first-order complexity theory. I am looking at a gradient method on a Riemannian manifold. A naive Euclidean analysis treats the manifold as a subset of a Euclidean space, takes a Lipschitz constant of the ambient gradient, and plugs this into standard Euclidean gradient-descent bounds. This gives the wrong dependence on the dimension. I want to understand the correct Riemannian framework and the relevant literature. Please do the following:
\begin{enumerate}
    \item Explain briefly why ambient Euclidean Lipschitz constants for $\nabla f$ can lead to suboptimal or dimension-misaligned complexity bounds for algorithms that are intrinsically Riemannian, and why the natural object is a Lipschitz constant for the Riemannian gradient (or for the pullback gradient under a retraction).
    \item List 3-6 key references (books or papers) on Riemannian gradient descent or related first-order methods with retractions and Lipschitz-continuous Riemannian gradients. For each reference, give a short summary (2-3 sentences) indicating:
    \begin{itemize}
        \item the type of manifold considered,
        \item the smoothness assumptions,
        \item the basic complexity statement.
    \end{itemize}
    \item From these references, extract a generic complexity theorem for Riemannian gradient descent with retractions and Lipschitz-continuous Riemannian gradient and list the main structural assumptions on the manifold, the retraction, and the function.
\end{enumerate}

\subsubsection{Prompts for Section \ref{sec: constant}}\label{appendix: coeffi}

You are a mathematical assistant with strong background in matrix manifolds and Riemannian optimization.

\textit{Setting.}
Work on the unitary group
\(
    \mathrm{U}(N) = \{ U \in \mathbb{C}^{N \times N} : U^\ast U = I \},
\)
equipped with the canonical Riemannian metric induced by the Frobenius inner product. Let $\|\cdot\|$ denote the Frobenius norm, and $T_x\mathrm{U}(N)$ the tangent space at $x \in \mathrm{U}(N)$.

We use the following length-5 product retraction [Equation \eqref{eq: 5-factor}],
where the real parameters $(a_1,a_2,b_1,b_2)$ are defined by \eqref{eq: 5 parameter}.
Here $H$ and $\psi_0$ are fixed Hermitian matrices, and $(x,y)$ are scalar coefficients associated with the tangent vector $\eta$ (you may assume a reasonable linear relation). The retraction at $U \in \mathrm{U}(N)$ in direction $\eta \in T_U\mathrm{U}(N)$ is defined by
\(
    \operatorname{Retr}_U(\eta) := \gamma_U(1;x_\eta,y_\eta)
\)
for the corresponding scalars $(x_\eta,y_\eta)$.

\textit{Goal.}
Find sharp global constants $\alpha$ and $\beta$ (ideally optimal, or close to optimal) such that, for all $U \in \mathrm{U}(N)$ and all tangent vectors $\eta \in T_U\mathrm{U}(N)$ in a suitable step-size range,
\(
   \|\operatorname{Retr}_U(\eta) - U\| \le \alpha \|\eta\|, \quad
\|\operatorname{Retr}_U(\eta) - U - \eta\| \le \beta \|\eta\|^2.
\)

\textit{Requirement.}
You must proceed in the following \emph{order}:
\begin{enumerate}
    \item {propose candidate constants and simple numerical tests.}
    \begin{itemize}
        \item Propose reasonable candidate constants $(\alpha,\beta)$, with a brief explanation.
        \item Then design and \emph{explicitly perform} simple numerical experiments that you would run to test these constants.
        \item Based on these tests, explain how you would confirm or adjust your candidate $(\alpha,\beta)$ (for example, by taking them slightly larger).
    \end{itemize}
    \item {give a theoretical justification.}
    \begin{itemize}
        \item Using the explicit product form of $\gamma_U(t;x,y)$, derive theoretical bounds that prove the two inequalities above with your final choice of constants (or with slightly relaxed constants if needed).
        \item Keep the argument as simple and direct as possible, using standard matrix inequalities and expansions, and make the dependence on $N$ clear if it appears.
    \end{itemize}
    \item  {comment briefly on tightness.}
    Explain whether the constants you obtain seem tight or near-tight, using both your reasoning and the (hypothetical) numerical evidence.
\end{enumerate}

\subsubsection{Prompts for Section \ref{sec: refining the theorem}}\label{appendix: refining the theorem}

You are an expert in both quantum algorithms and mathematical optimization. We are analyzing the convergence of a Riemannian gradient descent algorithm with retractions on $\mathrm{U}(N)$. The objective function is:
\(
f(U) = \operatorname{Tr}(H U \psi_0 U^\dagger)
\)
where $H$ and $\psi_0$ are Hermitian matrices. The manifold is equipped with the canonical metric induced by the Frobenius inner product.

\textit{Current Baseline Result}
We have successfully established a complexity bound. The current theorem is as follows: [Statement of Theorem \ref{thm-bese-complexity}].

\textit{Request for Deeper Theoretical Analysis}
The result above gives a sublinear rate ($O(\varepsilon^{-2})$), which applies to generic smooth non-convex functions. However, our objective function $f(U)$ has a very specific algebraic structure (a linear trace function on the unitary group). I suspect that the baseline analysis is too conservative because it ignores this specific structure. Please perform a deeper theoretical analysis to see if we can improve this bound:
\begin{enumerate}
    \item {Structural Analysis:} Analyze the algebraic relationship between the norm of the Riemannian gradient, $\|\operatorname{grad} f(U)\|$, and the function value gap. Does the specific form of $f(U)$ imply a stronger geometric property than simple smoothness?
    \item {Improved Convergence:} If such a relationship exists, does it allow us to prove a faster convergence rate (e.g., linear convergence instead of sublinear) under the same algorithmic settings?
    \item {Refined Theorem:} If you find a stronger bound, please formulate a new theorem that reflects this improved rate and the corresponding iteration count $T$.
\end{enumerate}

\section{Separate case study: composite \texorpdfstring{\((L_0,L_1)\)}{(L0,L1)}-smooth convex optimization}\label{appendix:l0l1-case}

This appendix presents a third case study on composite convex optimization under \((L_0,L_1)\)-smoothness. The prompt--response record and the complete human-finalized proof document are provided in the public repository at \url{https://github.com/optsuite/MathResearchPrompts}. We organize the discussion into four research-process subsections and summarize the final mathematical conclusion here. The LLM makes substantive contributions to problem scoping and to the development of the recentering, adaptive-acceleration, and hard-instance mechanisms. We determine the final statements and oracle scope, complete the derivations, and verify the proof.

We study the composite optimization problem \(\min_{x\in\mathbb R^d}F(x):=f(x)+h(x)\), where \(f\) is convex and differentiable, \(h\) is proper, closed, convex, and proximable, \(L_0>0\), \(L_1\geq0\), and \(\|\nabla^2 f(x)\|\leq L_0+L_1\|\nabla f(x)\|\). Assume that the solution set \(X^\star\) is nonempty, \(\operatorname{dist}(x_0,X^\star)\leq R\), and there are \(x^\star\in X^\star\) and \(v^\star\in\partial h(x^\star)\) satisfying \(\nabla f(x^\star)+v^\star=0\) and \(\|v^\star\|\leq G_\star\). Set \(L_\star:=L_0+L_1G_\star\).
The motivating smooth results have a small-\(\varepsilon\) leading term controlled by \(L_0\), because \(h\equiv0\) implies \(\nabla f(x^\star)=0\) \cite{gorbunov2024l0l1,tyurin2025near}. In the composite problem, \(\nabla f(x^\star)=-v^\star\) can be nonzero, so the effective curvature at optimality can instead be \(L_0+L_1G_\star\). We consider following three questions:
{
\begin{enumerate}
    \item Can a proximal method attain the accelerated leading order \(O(R\sqrt{L_\star/\varepsilon})\) with only lower-order warm-start terms?
    \item Can one construct a matching lower bound without allowing the proximal oracle to reveal the embedded hard smooth instance?
    \item Can an accelerated proximal method attain the upper bound while using only \(f(x)\), \(\nabla f(x)\), and \(\operatorname{prox}_{\lambda h}(x)\)?
\end{enumerate}}

The case therefore begins with a genuine theorem-discovery task. Neither the algorithm nor the lower-bound construction is specified by the research problem.

\begin{table}[htbp]
\centering
\scriptsize
\renewcommand{\arraystretch}{1.14}
\manualtablecaption{tab:l0l1-subsections}{Overview of the four research-process subsections in the separate, self-contained composite-optimization case. The LLM materially assists in developing the research route and key mechanisms. We complete and verify the final statements and proofs.}
\begin{tabularx}{0.98\linewidth}{>{\raggedright\arraybackslash}p{0.16\linewidth}>{\raggedright\arraybackslash}p{0.25\linewidth}>{\raggedright\arraybackslash}p{0.27\linewidth}X}
\hline
Subsection & AI task & Human verification & Resulting output \\
\hline
Problem scoping & Separate the composite problem from the known smooth case and identify the upper- and lower-bound obstacles. & Check the KKT interpretation, the effective scale, and which targets remain conjectural. & The scale \(L_0+L_1G_\star\) and a two-sided research roadmap. \\
Structural discovery & Search for an equivalent decomposition and inequalities that expose the correct curvature. & Verify the centered smoothness and gradient-gap relations. & Affine recentering, generalized cocoercivity, and a centered gradient bound. \\
Upper-bound development & Propose an executable accelerated proximal method with adaptive step sizes and a potential argument. & Check and fix the step-size rule, warm-start condition, Lyapunov recursion, and logarithmic terms. & An adaptive accelerated scheme with a two-phase complexity analysis. \\
Lower-bound discovery & Embed a smooth hard instance while realizing a \(G_\star\)-sized optimal certificate. & Check global smoothness, proximal-oracle independence, and the exact deterministic scope. & An orthogonal absolute-value lift and a span-respecting transcript reduction. \\
\hline
\end{tabularx}
\end{table}

\subsection{Problem scoping}\label{appendix:l0l1-subsection1}

The first request asks why the smooth and composite problems differ at an optimum, which scale should govern the small-accuracy regime, and what is required for matching upper and lower bounds. The model identifies the central obstruction: in the smooth case \(\nabla f(x^\star)=0\), whereas in the composite case \(\nabla f(x^\star)=-v^\star\) may have norm as large as \(G_\star\). It therefore proposes \(L_\star:=L_0+L_1G_\star\) as the effective curvature scale and separates the task into an accelerated upper-bound route and an oracle-preserving lower-bound route.

\textbf{Human disposition.} We verify the KKT interpretation and retain the proposed scale and two-sided roadmap. We reformulate the target as a near-optimal upper bound with leading term \(R\sqrt{L_\star/\varepsilon}\) and explicitly allow lower-order logarithmic terms, rather than assume an accuracy-independent localization cost. We also require the lower-bound construction to maintain a nonzero smooth-gradient component at an optimum while revealing no information about the embedded smooth hard instance through proximal queries. These checks turn the exploratory response into precise obligations for the remaining subsections.

\subsection{Structural discovery}\label{appendix:l0l1-subsection2}

The second request asks the model to search for a decomposition that removes the obstruction \(\nabla f(x^\star)\neq0\) without changing the executable forward--backward step, and for inequalities that connect the centered gradient to the objective gap. The model proposes an affine recentering and a generalized-cocoercivity route. These mechanisms are central to the final proof once we supply and check the exact identities.

\textbf{Human disposition.} Fix \(v^\star:=-\nabla f(x^\star)\in\partial h(x^\star)\) and define
\[
    \phi(x):=f(x)+\langle v^\star,x-x^\star\rangle,
    \qquad
    \psi(x):=h(x)-\langle v^\star,x-x^\star\rangle.
\]
Then \(F=\phi+\psi\), \(\nabla\phi(x^\star)=0\), and \(x^\star\) minimizes both centered components. With \(L_\star:=L_0+L_1G_\star\), the smooth component \(\phi\) is \((L_\star,L_1)\)-smooth. Although \(v^\star\) is unknown to the algorithm, it cancels from the complete proximal-gradient step,
{
\[
    \operatorname{prox}_{\lambda\psi}
    \bigl(y-\lambda\nabla\phi(y)\bigr)
    =
    \operatorname{prox}_{\lambda h}
    \bigl(y-\lambda\nabla f(y)\bigr).
\]}
We combine this identity with generalized cocoercivity \cite{tyurin2025near} to derive a centered gradient--gap bound. For
\[
    q(s):=\frac{s^2}{2(L_\star+4L_1s)},
\]
the finalized argument gives \(\phi(x)-\phi(x^\star)\geq q(\|\nabla\phi(x)\|)\), and hence \(\|\nabla\phi(x)\|\leq q^{-1}(F(x)-F^\star)\). This relation supplies the adaptive curvature control used in Subsection~\ref{appendix:l0l1-subsection3}.

\subsection{Upper-bound development and human proof reconstruction}\label{appendix:l0l1-subsection3}

The third request asks for an executable accelerated proximal method, an invariant that controls extrapolated iterates, and a rigorous rate. The model proposes adaptive step sizes tied to the centered gradient scale and a Lyapunov recursion. We retain this core route but replace the tentative localization claim in the interaction with an explicit step-size rule and a complete two-phase analysis.

\textbf{Human disposition.} Starting from \(y_0=u_0=x_0\), a radius bound \(R\geq\|x_0-x^\star\|\), and a warm-start scale \(\Gamma_0>0\), the finalized method uses
\[
    \gamma_k
    :=
    \frac{1}{L_\star+4L_1q^{-1}(\Gamma_kR^2)},
    \qquad
    \alpha_k:=\sqrt{\gamma_k\Gamma_k},
    \qquad
    \Gamma_{k+1}:=\frac{\Gamma_k}{1+\alpha_k}.
\]
It combines an extrapolated proximal-gradient point with an auxiliary update defined by a subgradient of \(F\). The central potential is
\[
    V_k
    :=
    F(y_k)-F^\star
    +\frac{\Gamma_k}{2}\|u_k-x^\star\|^2.
\]
Under the stated warm-start condition, we prove \(V_{k+1}\leq V_k/(1+\alpha_k)\) and \(V_k\leq\Gamma_kR^2\). The large-error phase decreases \(\Gamma_kR^2\) geometrically, while the small-error phase applies an accelerated auxiliary-sequence bound. This yields
{
\[
    N
    =
    O\!\left(
        R\sqrt{\frac{L_\star}{\varepsilon}}
        +\log\!\left(2+\frac{\Gamma_0}{L_\star}\right)
        +(L_1R+1)\log\!\left(2+\frac{\Gamma_0R^2}{\varepsilon}\right)
    \right).
\]}
The first term is the accelerated leading order, and the remaining terms record the warm-start and generalized-smoothness overhead. The displayed scheme assumes valid bounds \(R\) and \(G_\star\). If they are unavailable, an additional doubling or backtracking wrapper is required.

\subsection{Lower-bound discovery and correction}\label{appendix:l0l1-subsection4}

The final request asks how to embed a classical \(L_\star\)-smooth convex hard instance into the composite class while retaining an optimal certificate of size \(G_\star\), without allowing the proximal oracle to reveal the hard block. The model compares same-space and orthogonal-coordinate embeddings and identifies the latter as the robust route. We reject a same-space affine tilt because the hard gradient can cancel the added \(G_\star\)-component away from the optimum.

\textbf{Human disposition.} Let \(g:\mathbb R^{d-1}\to\mathbb R\) be a classical \(L_\star\)-smooth convex hard instance and write \(x=(z,r)\). The finalized construction is
{
\[
    f(z,r):=g(z)-G_\star r,
    \qquad
    h(z,r):=G_\star|r|.
\]}
Then \(F(z,r)=g(z)+G_\star(|r|-r)\), \(\|\nabla f(z,r)\|\geq G_\star\), and
\[
    \|\nabla^2f(z,r)\|
    \leq L_\star
    =L_0+L_1G_\star
    \leq L_0+L_1\|\nabla f(z,r)\|.
\]
At \(x^\star=(z^\star,0)\), a subgradient of \(h\) cancels \(\nabla f(x^\star)\) and has norm \(G_\star\). The proximal map is the identity on the \(z\)-coordinate and soft-thresholding on the \(r\)-coordinate, so it reveals no additional information about \(g\). Projecting the composite transcript onto the \(z\)-block therefore transfers the classical smooth lower bound \cite{bubeck2015convex}. We restrict the finalized claim to span-respecting deterministic proximal first-order methods whose final output is one of the queried points. No unconditional lower bound is claimed here for unrestricted deterministic outputs or randomized methods.

\subsection{Final conclusion and detailed proof record}\label{appendix:l0l1-final}

The four research-process subsections above summarize the development of the research direction and core mechanisms. The following result records the final mathematical conclusion, which we establish and verify. We state one combined summary theorem here rather than reproducing every supporting statement. The complete algorithm, five supporting lemmas, two theorem proofs, and exact constants appear in \path{materials/l0l1/l0l1.pdf}.

{
\begin{theorem}[Composite \((L_0,L_1)\)-smooth complexity]\label{thm:l0l1-final}
Let \(F=f+h\), where \(L_0>0\), \(L_1\geq0\), and \(f\) is convex, twice differentiable, and satisfies
\[
    \|\nabla^2 f(x)\|
    \leq
    L_0+L_1\|\nabla f(x)\|,
\]
and \(h\) is proper, closed, convex, and proximable. Assume \(G_\star\geq0\) and that there are \(x^\star\in\operatorname*{argmin}F\) and \(v^\star\in\partial h(x^\star)\) such that
\[
    \nabla f(x^\star)+v^\star=0,
    \qquad
    \|x_0-x^\star\|\leq R,
    \qquad
    \|v^\star\|\leq G_\star,
\]
and set \(L_\star:=L_0+L_1G_\star\). If \(R>0\) and
\[
    \Gamma_0
    \geq
    \frac{2(F(x_0)-F^\star)}{R^2},
\]
then the adaptive accelerated proximal-gradient method summarized in Subsection~\ref{appendix:l0l1-subsection3} reaches \(F(y_N)-F^\star\leq\varepsilon\) in
\[
    N
    =
    O\!\left(
        R\sqrt{\frac{L_\star}{\varepsilon}}
        +\log\!\left(2+\frac{\Gamma_0}{L_\star}\right)
        +(L_1R+1)\log\!\left(2+\frac{\Gamma_0R^2}{\varepsilon}\right)
    \right)
\]
iterations. Conversely, for every \(k\geq1\) and \(d\geq2k+2\), there is an instance in this class such that every span-respecting deterministic proximal first-order method whose final output is one of its queried points and that uses at most \(k\) oracle calls returns \(x_k\) with
\[
    F(x_k)-F^\star
    \geq
    \frac{3L_\star R^2}{32(k+1)^2}.
\]
Thus \(R\sqrt{L_\star/\varepsilon}\) is the tight leading-order dependence in this oracle model, while the upper bound retains the displayed lower-order logarithmic terms.
\end{theorem}}

Overall, in this example, the LLM is used to help formulate the composite-optimization problem, identify the effective curvature scale \(L_0+L_1G_\star\), explore affine recentering and an adaptive accelerated route for the upper bound, and develop an oracle-preserving hard-instance construction for the lower bound. The interaction illustrates how these ideas are progressively refined into a coherent route from problem scoping to matching leading-order upper and lower bounds. Further technical details and the complete proof are provided in \url{https://github.com/optsuite/MathResearchPrompts}.

\bibliographystyle{plain}
\bibliography{references}
\end{document}